\documentclass[
 reprint,
 amsmath,amssymb,
 aps,
]{revtex4-2}
\usepackage{graphicx}
\usepackage{dcolumn}
\usepackage{bm}
\usepackage{booktabs}     
\usepackage{dcolumn}      
\newcolumntype{.}{D{.}{.}{4}} 
\usepackage{graphicx}
\usepackage{amsmath}  
\usepackage{hyphenat} 

\usepackage[utf8]{inputenc}
\usepackage{graphicx}
\usepackage{dcolumn}
\usepackage{bm}
\usepackage{float}
\usepackage{amsmath}
\usepackage{graphicx,subfigure,epsfig}
\usepackage{hyperref}
\hypersetup{
    colorlinks=true,
    citecolor=blue,
    linkcolor=blue,
    filecolor=magenta,
    urlcolor=cyan,}
\makeatletter
\def\thetable{\@arabic\c@table} 
\makeatother

\begin{document}

\preprint{APS/123-QED}

\title{Optical Appearance and Ringdown of Black Holes in a Kalb–Ramond Field Coupled to Perfect Fluid Dark Matter}

\author{Qi-Qi Liang$^{1}$}
\author{Zi-Qiang Cai$^{1}$}
\author{Dong Liu$^{2}$}
\author{Zheng-Wen Long$^{1}$}
\email[Corresponding author: ]{zwlong@gzu.edu.cn}

\affiliation{$^{1}$College of Physics, Guizhou University, Guiyang, 550025, China}
\affiliation{$^{2}$Department of Physics, Guizhou Minzu University, Guiyang, 550025, China}

\begin{abstract}

This paper investigates the optical and dynamical properties of a static spherically symmetric black hole in the presence of a Kalb--Ramond (KR) field coupled to perfect fluid dark matter (PFDM). We analyze the effects of the Lorentz-violating parameter $\alpha$ and the dark matter parameter $\lambda$ on photon trajectories and their observational signatures in the strong-gravity regime. Furthermore, we study the quasinormal mode spectrum under scalar, electromagnetic, and gravitational perturbations, examining how the model parameters influence the characteristic oscillation frequencies and damping rates. In particular, the interplay between the effective potential structure and perturbative dynamics is clarified, and it is found that, within the validity of the eikonal approximation, the quasinormal modes of the black hole considered here exhibit good agreement with the properties of null geodesics. Our results show that the model parameters significantly affect both the optical appearance of the black hole and the dynamical features of the ringdown phase, providing potential observational constraints on Lorentz-violating effects and dark matter environments in strong-field regimes.

\end{abstract}

\maketitle

\section{Introduction}
\label{sec:1}
In recent years, breakthrough progress in gravitational-wave detections and black hole imaging observations has opened new avenues for probing the spacetime structure in the strong-gravity regime. In 2015, the LIGO Scientific Collaboration reported the first direct detection of gravitational waves generated by the merger of a binary black hole system~\cite{LIGOScientific:2016aoc}.
In the ringdown phase after a black hole merger, or in the late-time evolution of a perturbed black hole, the system returns to equilibrium through damped oscillations described by quasinormal modes~\cite{Berti:2009kk,Kokkotas:1999bd,Konoplya:2011qq,RibesMetidieri:2025lxr}. As these modes depend solely on the spacetime geometry of the black hole, their frequencies and damping times provide an important tool for probing black hole structure and testing theories of gravity.
In addition, in 2019 the Event Horizon Telescope Collaboration reported the first image of the supermassive black hole M87* ~\cite{EventHorizonTelescope:2019ths}, revealing the black hole shadow and the surrounding photon ring structure~\cite{Johnson:2019ljv}. Photon rings arise from photons that execute multiple orbits near unstable photon trajectories, and their geometric features encode valuable information about the spacetime geometry near the black hole~\cite{Falcke:1999pj,Luminet:1979nyg,Urso:2025gos,Desire:2024mzp}. Consequently, optical appearance  and quasinormal modes serve as important probes for studying black hole physics and testing gravitational theories.

In realistic astrophysical environments, black holes are rarely isolated systems but are instead embedded in complex backgrounds composed of various matter fields and cosmic media~\cite{Perivolaropoulos:2021jda,Fernandes:2025osu,Narayan:2008bv}. Among these components, dark matter is believed to constitute the dominant fraction of the matter content of the Universe and plays a crucial role in galaxy evolution and the formation of large-scale structures~\cite{Bertone:2004pz,Roszkowski:2017nbc,Planck:2018vyg}. Observational evidence indicates that the presence of dark matter can significantly modify the gravitational distribution and dynamical behavior of astrophysical systems~\cite{Rubin:1970zza,Peebles:1982ff}. To describe the macroscopic properties of dark matter in gravitational fields, a variety of models have been proposed~\cite{Blumenthal:1984bp,Irsic:2017ixq,Ludlow:2016ifl,Navarro:1996gj}. Among them, the perfect fluid dark matter model has been widely employed because it can effectively reproduce galactic rotation curves and capture the global features of dark matter halos~\cite{Bharadwaj:2003iw}. Therefore, incorporating dark matter backgrounds into black hole models is of considerable importance for understanding their influence on spacetime geometry, photon propagation, and gravitational-wave signals.

On the other hand, general relativity may not represent the ultimate theory of gravity under extreme high-energy or strong-curvature conditions, motivating the exploration of extended gravitational models with additional degrees of freedom~\cite{Will:2014kxa}. Various modified theories, such as scalar-tensor models~\cite{Damour:1993hw,Sotiriou:2011dz,Herdeiro:2014goa}, higher-curvature corrections (e.g., Gauss-Bonnet or $f(R)$ gravity)~\cite{Sotiriou:2008rp,Boulware:1985wk,Lyu:2022gdr}, and non-Abelian gauge field theories~\cite{Balakin:2006gv,Balakin:2015gpq,Lutfuoglu:2025ljm,Liu:2019pov,Hod:2025cfm}, can give rise to black hole spacetimes that deviate from the standard solutions of general relativity, thereby providing new avenues for testing gravitational physics.

Within these extended frameworks, the Kalb-Ramond field, originating from the low-energy effective action of string theory, is a rank-2 antisymmetric tensor field~\cite{Kalb:1974yc}. A nonvanishing background configuration of this field may lead to spontaneous Lorentz symmetry breaking and modify the effective gravitational field equations, resulting in novel black hole solutions~\cite{Sucu:2025lqa,Kumar:2020hgm,Mangut:2025gie,Duan:2023gng,Shi:2025rfq,Ditta:2024lnb,Filho:2023ycx,AraujoFilho:2025jcu,
AraujoFilho:2024ctw}. When such a field is present around a black hole, the thermodynamic properties, particle trajectories, and propagation of perturbations may differ significantly from those in the standard scenario.

Recent studies have further considered the coexistence of the Kalb-Ramond field and perfect fluid dark matter, showing that their combined effects can substantially alter the geometric structure and dynamical properties of black hole spacetimes~\cite{Jha:2025uie,Shodikulov:2025xax,Rahmatov:2025gpk,Jumaniyozov:2025dyy}. However, a comprehensive investigation of the observational appearance, photon-ring characteristics, and quasinormal mode behavior during the ringdown phase in this background is still lacking. In fact, the photon sphere not only determines the morphology of the black hole shadow and photon rings but also governs the quasinormal mode spectrum in the eikonal limit. Therefore, a joint analysis of optical features and ringdown properties can provide a more complete understanding of how external fields modify observable signals from black holes.

Motivated by the above considerations, we investigate the optical appearance and quasinormal modes of static, spherically symmetric black holes in the background of a Kalb--Ramond field coupled to perfect fluid dark matter. In Sec.~\ref{sec:2}, we introduce the theoretical framework of the Kalb-Ramond field coupled to perfect fluid dark matter and present the corresponding static spherically symmetric black hole solution. In Sec.~\ref{sec:3}, we investigate photon motion in this spacetime, analyze the photon-ring structure, and construct a thin accretion disk model to obtain the observed image. In Sec.~\ref{sec:qnm}, we study the quasinormal mode spectrum generated by perturbations of different spin fields, discuss the influence of model parameters on the ringdown characteristics, and verify the correspondence between the QNM spectrum and the orbital angular velocity as well as the Lyapunov exponent of the photon sphere. Finally, Sec.~\ref{sec:5} summarizes the main results of this work. All quantities in this paper are expressed in geometric units with $M=G=c=1$.

\section{Black hole solution in KR field with PFDM}
\label{sec:2}
In this section, we summarize the black hole solution reported in Ref.~\cite{Jumaniyozov:2025dyy}, which corresponds to a static and spherically symmetric spacetime influenced by both a background Kalb--Ramond (KR) field and a surrounding perfect fluid dark matter distribution. The dynamics of the system are governed by Einstein gravity with a nonminimal coupling to the KR field, supplemented by a perfect fluid dark matter component. The total action can be written as the sum of three parts,

\begin{equation}
S = S_{GR} + S_{KR} + S_{PFDM},
\end{equation}

where $S_{GR}$ denotes the Einstein--Hilbert gravitational action,
$S_{KR}$ represents the contribution from the Kalb--Ramond field,
$S_{PFDM}$ corresponds to the contribution of perfect fluid dark matter. 
The Kalb--Ramond field is described by a rank-2 antisymmetric tensor $B_{\mu\nu}$. 
Its corresponding field strength tensor is defined as 
$H_{\mu\nu\rho} = \partial_{\mu} B_{\nu\rho} + \partial_{\nu} B_{\rho\mu} + \partial_{\rho} B_{\mu\nu}$. 
The energy-momentum tensor of the PFDM is modeled as a perfect fluid, 
$T^{PFDM}_{\mu\nu} = (\rho + p) u_\mu u_\nu + p\, g_{\mu\nu}$, 
where $\rho$ and $p$ denote the energy density and pressure, respectively, and $u_\mu$ is the four-velocity of the fluid.

By varying the total action with respect to the metric $g_{\mu\nu}$, one obtains the gravitational field equations:
\begin{equation}
R_{\mu\nu} - \frac{1}{2} g_{\mu\nu} R = T^{KR}_{\mu\nu} + T^{PFDM}_{\mu\nu}
\end{equation}

The field equations follow from varying the above action with respect to the metric and the KR field. Imposing staticity and spherical symmetry, one arrives at a class of black hole solutions characterized by the line element:
\begin{equation}
ds^{2} = -f(r)\,dt^{2} + \frac{1}{f(r)}\,dr^{2} + r^{2}(d\theta^{2} + \sin^{2}\theta\, d\varphi^{2})
\label{equ:2}
\end{equation}
with the metric function expressed as:
\begin{equation}
f(r) = \frac{1}{1-\alpha} - \frac{2M}{r} + \frac{\lambda}{r}\ln\!\left(\frac{r}{|\lambda|}\right)
\label{equ:3}
\end{equation}

The metric function $f(r)$ depends on the mass parameter $M$, the Lorentz-violation parameter $\alpha$ associated with the nontrivial background configuration of the Kalb–Ramond field, and the perfect fluid dark matter parameter $\lambda$.The event horizon is determined by the largest root of the equation $f(r)=0$. In the limit $\alpha=0$ and $\lambda=0$, the solution reduces to the Schwarzschild black hole, which possesses a physical singularity at $r=0$.

To ensure the physical viability of the model, the quasiperiodic oscillations frequencies of test particles were analyzed in Ref.~\cite{Jumaniyozov:2025dyy}, where the model parameters were constrained by considering different types of black hole systems. In the following discussion, we adopt these constraint ranges as a reference. In particular, the parameter $\alpha$ is restricted to the interval $(0.05,\,0.45)$, while $\lambda$ is taken within the range $(0.005,\,0.045)$.

\section{Black hole shadow, photon ring, thin accretion disk}
\label{sec:3}
\subsection{Shadow and photon rings}

To study the black hole shadow, it is essential to investigate the motion of photons in the vicinity of black holes. In such regions, gravity becomes extremely intense, and spacetime curvature significantly affects photon propagation, resulting in distinct trajectories for different photons. These varied orbital structures of photons directly determine the geometric shapes of the black hole shadow and the photon ring. Since the black hole considered here is spherically symmetric, the photon motion is confined to a single plane. Without loss of generality, we restrict our analysis to the equatorial plane $(\theta = \pi/2)$. The photon trajectory is then governed by the following radial equation:

\begin{equation}
\left(\frac{dr}{d\phi}\right)^2
= r^4 \left( \frac{1}{b^2} - \frac{f(r)}{r^2} \right)
\label{equ:7}
\end{equation}

Here, $\phi$ denotes the azimuthal angle on the equatorial plane, and $b = L/E$ represents the impact parameter, where $E$ and $L$ are the conserved energy and angular momentum of the photon, respectively. To facilitate the analysis of the photon’s radial motion, the expression in parentheses $\left( \frac{1}{b^2} - \frac{f(r)}{r^2} \right)$, is defined as the effective potential.

The orbital properties of photons are determined by the impact parameter $b$. The radius of the photon sphere $r_{\mathrm{ph}}$, formed by bounded photon orbits, 
can be obtained from the extremum conditions of the effective potential. Specifically, at $r = r_{\mathrm{ph}}$,The effective potential and its derivative with respect to the radial coordinate are simultaneously equal to zero.

\begin{equation}
\left( \frac{1}{b^2} - \frac{f(r_{\mathrm{ph}})}{r_{\mathrm{ph}}^2} \right)= 0
\label{equ:8}
\end{equation}
\begin{equation}
\left.\frac{d\left( \frac{1}{b^2} - \frac{f(r)}{r^2} \right)}{dr}\right|_{r = r_{\mathrm{ph}}} = 0
\label{equ:9}
\end{equation}

The photon sphere radius $r_{\mathrm{ph}}$ is obtained from Eq.~\ref{equ:9}. Substituting $r_{\mathrm{ph}}$ into Eq.~\ref{equ:8} yields the critical impact parameter $b_c$ corresponding to the photon sphere. This parameter represents the critical value for photons that can move along unstable 
circular orbits, and its expression is given by:

\begin{equation}
b_c = \frac{r_{\mathrm{ph}}}{\sqrt{f(r_{\mathrm{ph}})}}
\label{equ:10}
\end{equation}

To characterize the photon trajectories from spatial infinity to the vicinity of the black hole, we redefine the radial coordinate by introducing $u = 1/r$, which recasts the orbital equation into a form suitable for analyzing the bending of light:

\begin{equation}
G(u)=\left(\frac{du}{d\phi}\right)^2
= \frac{1}{b^2} - f\!\left(\frac{1}{u}\right) u^2
\label{equ:11}
\end{equation}

\begin{figure}[h!]
  \centering
  \includegraphics[width=\columnwidth]{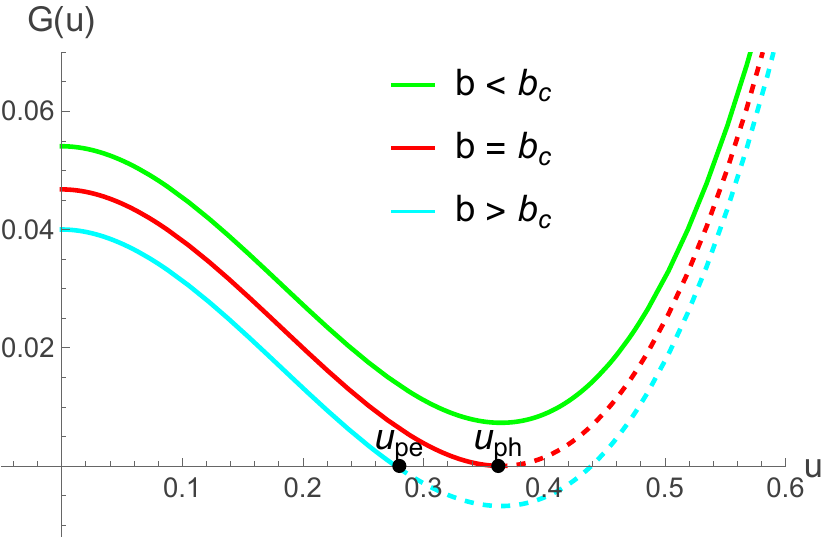}
  \caption{Plot of $G(u)$ for photon trajectories in the black hole spacetime when $\alpha = 0.05$ and $\lambda = 0.015$.}
  \label{fig:1}
\end{figure}

\begin{figure*}[t!]
  \includegraphics[width=0.48\textwidth]{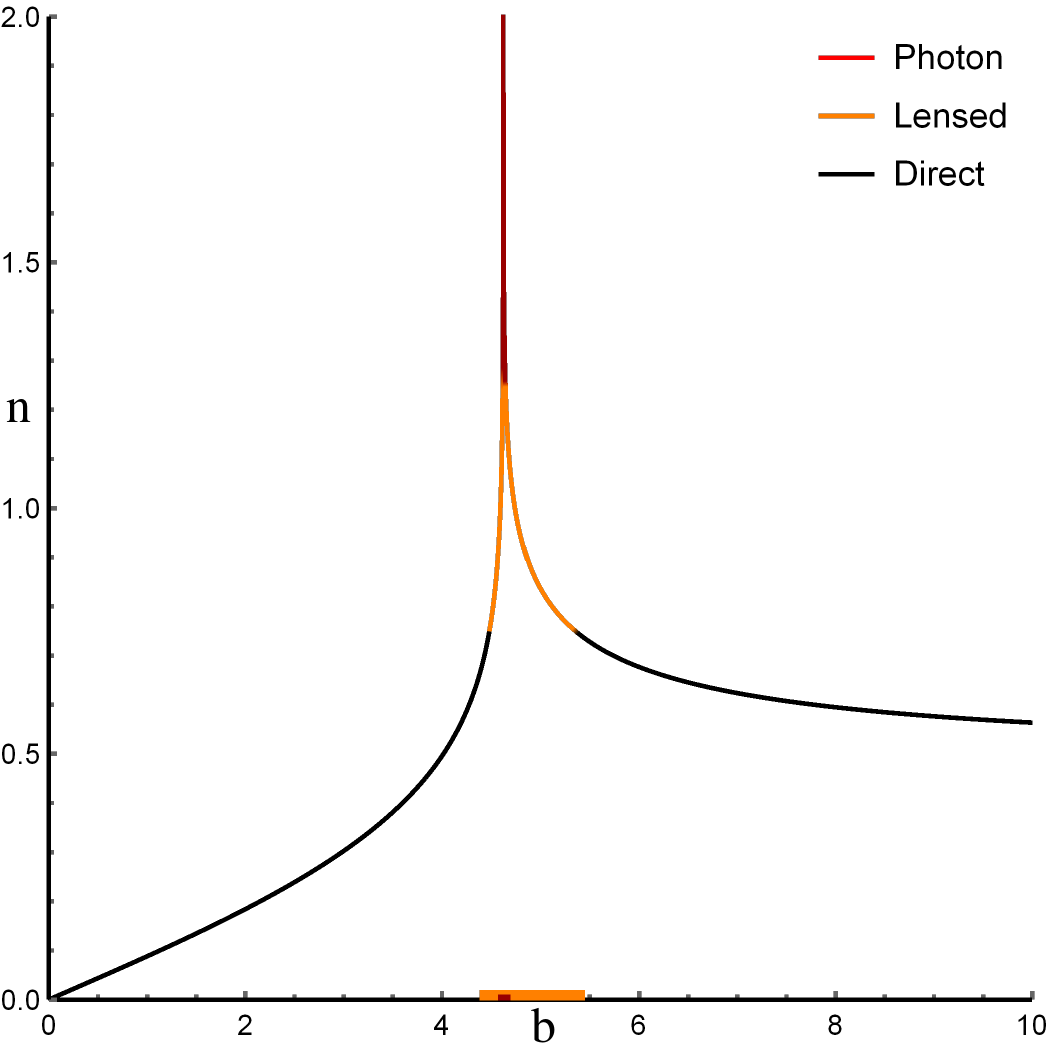}
    \includegraphics[width=0.48\textwidth]{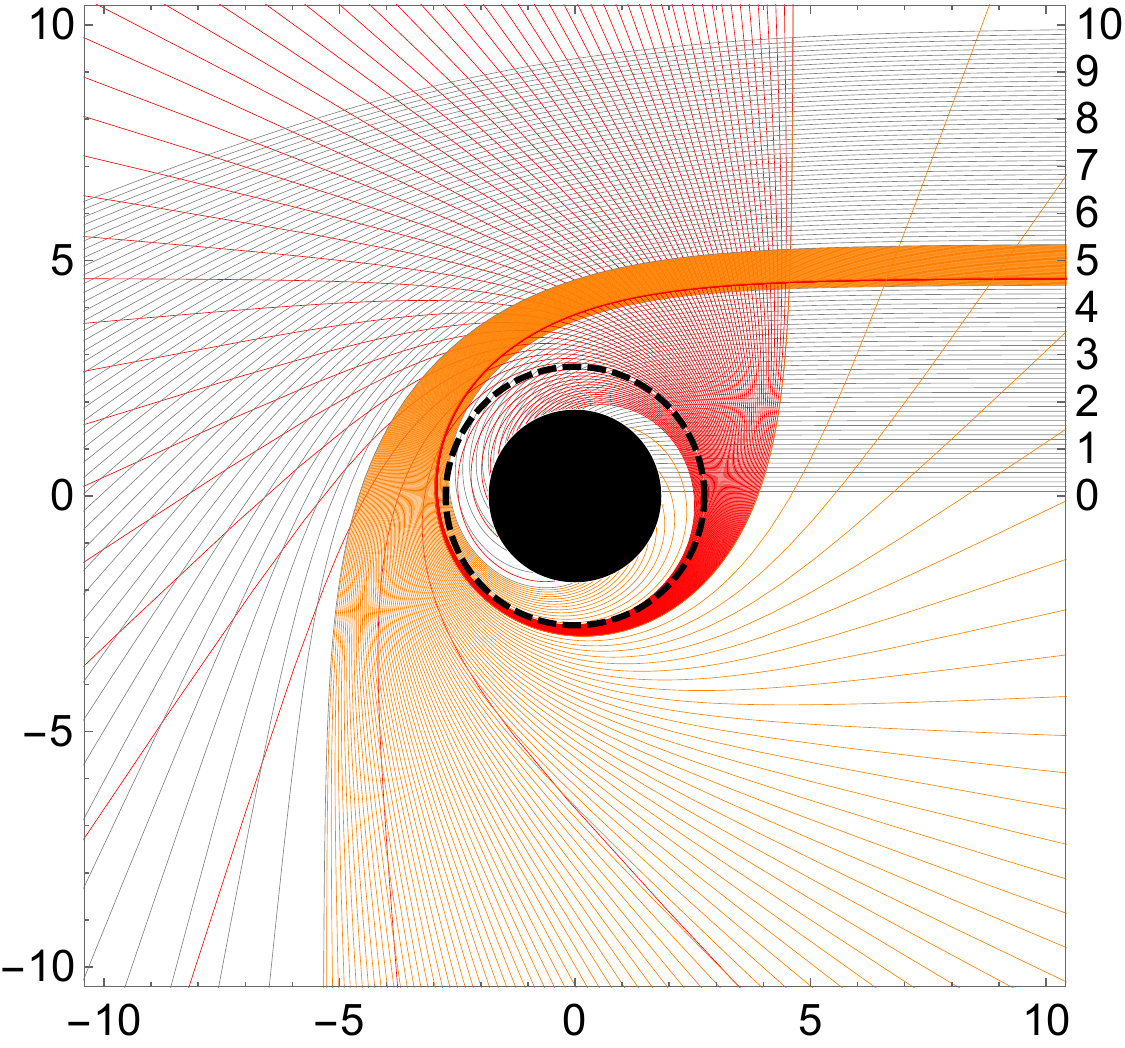}

\caption{ Relationship between the number of photon orbits $n$ and the impact parameter $b$ in the black hole spacetime when $\alpha = 0.05$ and $\lambda = 0.015$. The right panel shows the photon trajectories in polar coordinates $(r,\phi)$ for different impact parameters.}
\label{fig:2}     
\end{figure*}
As shown in Fig.~\ref{fig:1}, photons with different impact parameters propagate from spatial infinity toward the black hole, and their radial motion is constrained by the effective potential. The solid curves indicate the allowed regions accessible to photons, whereas the dashed curves represent the forbidden regions that photons cannot enter.When the impact parameter satisfies $b > b_{\rm c}$, the photon is incident from infinity and approaches the black hole. After reaching its minimum radial distance (corresponding to $u_{\rm pe}$ in the Fig.~\ref{fig:1}), it moves away along a symmetric trajectory and eventually returns to infinity. The deflection angle associated with this process is given by:

\begin{equation}
\varphi = 2 \int_{0}^{u_{\rm pe}} \frac{1}{\sqrt{G(u)}} \, du
\label{equ:13}
\end{equation}

When the impact parameter satisfies $b < b_c$, a photon incoming from infinity continues to move inward toward the black hole without developing a radial turning point. Consequently, it does not return to distant regions but instead crosses the event horizon and is ultimately absorbed by the black hole. For such photons, we consider only the motion outside the horizon, and the corresponding deflection angle is given by:

\begin{equation}
\varphi = \int_{0}^{u_{\rm h}} \frac{1}{\sqrt{G(u)}} \, du
\label{equ:12}
\end{equation}

In this case, $u_{\rm h}$ corresponds to the event horizon located at $r_{\rm h}$. When $b = b_{\rm c}$, the incoming photon follows a critical trajectory. After approaching the black hole from infinity, it neither escapes back to distant regions nor plunges directly into the black hole, but instead spirals around an unstable circular orbit at that radius (corresponding to $u_{\rm ph}$ in the Fig.~\ref{fig:1}). By expressing the photon’s angular displacement in terms of the number of revolutions $n$, defined as $n = \varphi / (2\pi)$, the function relating the number of revolutions $n$ to the impact parameter $b$ can be written as:

\begin{equation}
n(b) = \frac{i}{2} - \frac{1}{4}, \qquad i = 1, 2, 3, \cdots
\label{equ:14}
\end{equation}

The solutions of Eq.~\ref{equ:14} can be written as $b_{i}^{\pm}$, where $b_{i}^{-}$ corresponds to the solution smaller than $b_c$, and $b_{i}^{+}$ corresponds to the solution larger than $b_c$. These solutions can be used to distinguish the ranges of direct rays, lensed rays, and photon-ring rays, as detailed below~\cite{Gralla:2019xty}:

\begin{itemize}
  \item Direct: $  n < \tfrac{3}{4} \iff b \in (0, b_{2}^{-}) \cup (b_{2}^{+}, \infty) $;
  
  \item Lensing: $ \tfrac{3}{4} < n < \tfrac{5}{4} \iff b \in (b_{2}^{-}, b_{3}^{-}) \cup (b_{3}^{+}, b_{2}^{+}) $;
  
  \item Photon: $ n > \tfrac{5}{4} \iff b \in (b_{3}^{-}, b_{3}^{+}) $.
\end{itemize}

To intuitively demonstrate the motion trajectories of light rays under different impact parameters, the left panel of Fig.~\ref{fig:2} shows the corresponding relationship between the photon impact parameter $b$ and the number of orbital revolutions $n$. The right panel displays the photon trajectories in polar coordinates, where the central black region represents the black hole. As the impact parameter of the photon approaches the critical impact parameter $b_{\rm c}$, the number of orbital cycles $n$ increases. In this case, $b_{\rm c}$ corresponds to the radius of the black hole shadow as observed by a distant observer.

\begin{table*}[t!]
\centering
\caption{Values of various physical quantities of the black hole spacetime for different values of $\alpha$ and $\lambda$.}
\begin{tabular*}{\linewidth}{@{\extracolsep{\fill}} ccccccccccc @{}}
\toprule
$\alpha$ & $\lambda$ & $r_h$ & $r_{ph}$ & $b_c$ & $r_{\text{isco}}$ &$b_1^-$ &$b_2^-$ & $b_2^+$ & $b_3^-$ & $b_3^+$ \\
\midrule
0.05 & 0 & 1.9 & 2.85 & 4.81135 & 5.7& 2.69203& 4.66207 & 5.58054 & 4.80505 & 4.83529 \\
0.05 & 0.015 & 1.83153 & 2.74577 & 4.62341 & 5.50477& 2.59351& 4.4822 & 5.35137 & 4.61752 & 4.64583 \\
0.05 & 0.03 & 1.78357 & 2.67232 & 4.48754 & 5.37138& 2.52406& 4.35273 & 5.18286 & 4.48199 & 4.50872 \\
0.05 & 0.045 & 1.74366 & 2.61098 & 4.37219 & 5.26247& 2.46601& 4.2431 & 5.03841 & 4.36695 & 4.39226 \\
 \multicolumn{2}{c}{Schwarzschild} & 2  & 3  & 5.19615  & 6 & 2.8477& 5.01514  & 6.16757  & 5.18781  & 5.22794 \\
0 & 0.005 & 1.97012 & 2.95464 & 5.11326 & 5.91389 &2.80464 &4.93601 & 6.0636 & 5.10512 & 5.14428 \\
0.15 & 0.005 & 1.67529 & 2.51248 & 4.00871 & 5.02887& 2.34605& 3.91369 & 4.46328 & 4.00543 & 4.02112 \\
0.3 & 0.005 & 1.38033 & 2.07011 & 2.99735 & 4.14346 &1.88902 &2.95443 & 3.18409 & 2.99632 & 3.00122 \\
0.45 & 0.005 & 1.0852 & 1.62751 & 2.08881 & 3.25756 &1.43477 &2.07408 & 2.14839 & 2.0886 & 2.08962 \\
\bottomrule
\end{tabular*}
\label{tab:1} 
\end{table*}

To study thin accretion disks around black holes, it is necessary to introduce timelike geodesics. Since particles in the accretion disk undergo timelike motion in spacetime, the analysis method used for photons can be extended to obtain the radial motion of the particles as follows:

\begin{equation}
\left(\frac{dr}{d\phi}\right)^2
= r^4 \left( \frac{1}{b^2} - \frac{f(r)}{r^2} -\frac{f(r)}{L^2}\right)
\label{equ:15}
\end{equation}

In Eq.~\ref{equ:15}, by defining $\left( \frac{1}{b^2} - \frac{f(r)}{r^2} -\frac{f(r)}{L^2}\right)=U(r)$ as the effective potential, the radius of the innermost stable circular orbit $r_{\mathrm{isco}}$ can be obtained by simultaneously solving the following equations:

\begin{equation}
U(r) = \frac{dU(r)}{dr} = \left. \frac{d^2 U(r)}{dr^2} \right|_{r=r_{\text{isco}}} = 0
\label{equ:16}
\end{equation}

In the preceding section, we have discussed the motion of photons and particles in a static spherically symmetric black hole spacetime. For the black hole model considered in this work, the specific values of various physical quantities for different parameter choices are listed in Table~\ref{tab:1}. It can be seen that the model parameters have a significant influence on the spacetime structure in the vicinity of the black hole, thereby modifying observable quantities such as the photon-sphere radius, the critical impact parameter, and the shadow size. Specifically, for a fixed parameter $\alpha$, these physical quantities exhibit an overall decreasing trend as  $\lambda$increases; for a fixed $\lambda$, they also decrease as $\alpha$ increases.

\subsection{Thin Accretion Disk}
Black holes are typically surrounded by accreting matter, and the electromagnetic radiation detected from black hole systems primarily originates from the accretion disk. Therefore, studying photon orbits and black hole shadows alone is insufficient to fully characterize the properties of black holes. In this section, we investigate the thin accretion disk in the equatorial plane of the black hole spacetime considered in this work.

Considering the radiation from a thin accretion disk around a black hole, the light detected by a distant observer located at the north pole originates from the intersection points between the photon trajectories and the disk. Due to gravitational redshift, the observed frequency and the emitted frequency satisfy $\nu' = g \nu$, where $g$ is the redshift factor. According to Liouville’s theorem, the ratio of the specific intensity to the cube of the frequency remains invariant along a ray, which leads to the relation between the observed and emitted intensities $I^{\rm obs}_{\nu'} = g^3 I^{\rm em}_{\nu}$. After integrating over frequency, the total observed intensity becomes $I^{\rm obs} = g^4 I^{\rm em}(r)$. 

Moreover, because photons in a strong gravitational field may orbit the black hole multiple times and intersect the accretion disk repeatedly, the total observed intensity should be given by the sum of the contributions from all intersection points. Therefore, for a given impact parameter $b$, the total observed intensity can be expressed as:
\begin{equation}
I^{\rm obs}(b)=\sum_i g^4 I^{\rm em}\big|_{r=r_i(b)} 
\label{equ:17}
\end{equation}

The function $r_i(b)$, known as the transfer function, can be expressed in terms of the solution $u(b,\phi)$ of Eq.~\ref{equ:11} as follows:

\begin{equation}
r_i(b)=\frac{1}{u\!\left(\frac{(2i-1)\pi}{2},\, b\right)}, \quad i=1,2,3,\cdots
\label{equ:18}
\end{equation}

\begin{figure}[h!]
  \centering
  \includegraphics[width=\columnwidth]{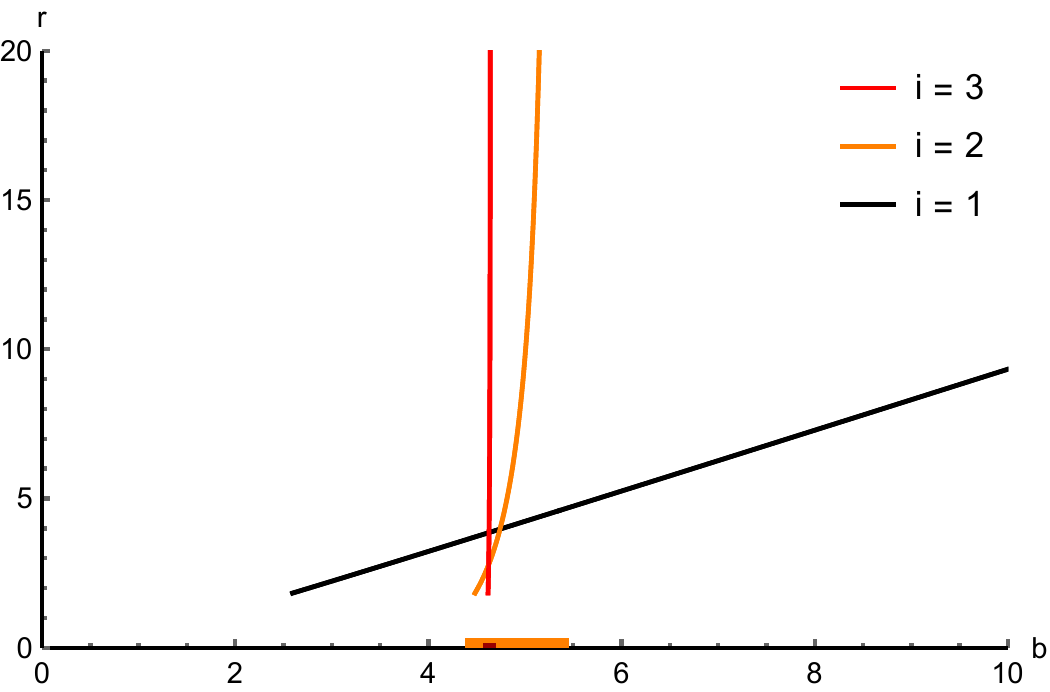}
  \caption{The first three transfer functions of a thin accretion disk around the black hole spacetime when  $\alpha = 0.05$ and $\lambda = 0.015$.}
  \label{fig:3}
\end{figure}

\begin{figure*}[t!]
\centering
\begin{minipage}{0.34\textwidth}
    \includegraphics[width=\linewidth]{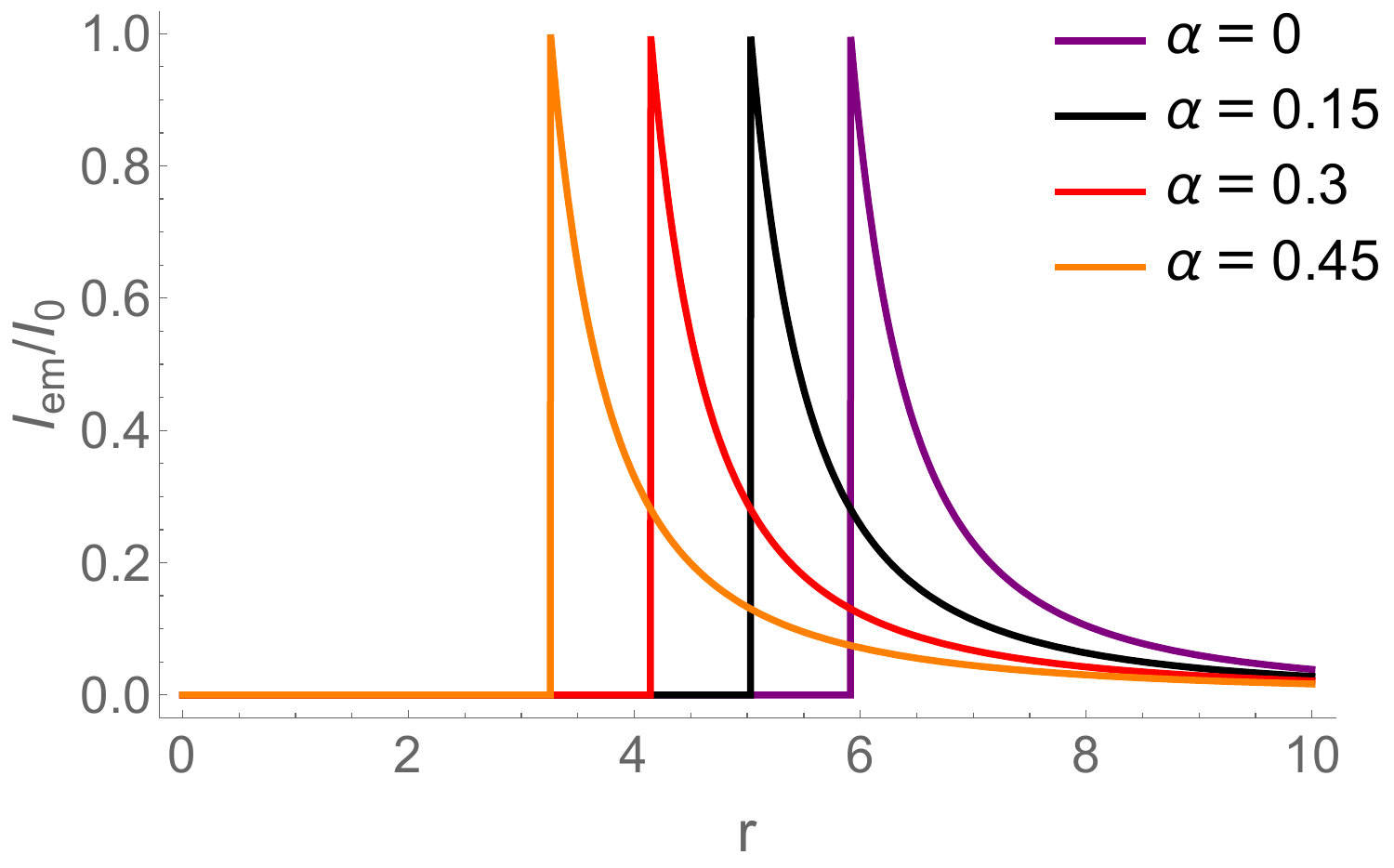}
\end{minipage}
\begin{minipage}{0.34\textwidth}
    \includegraphics[width=\linewidth]{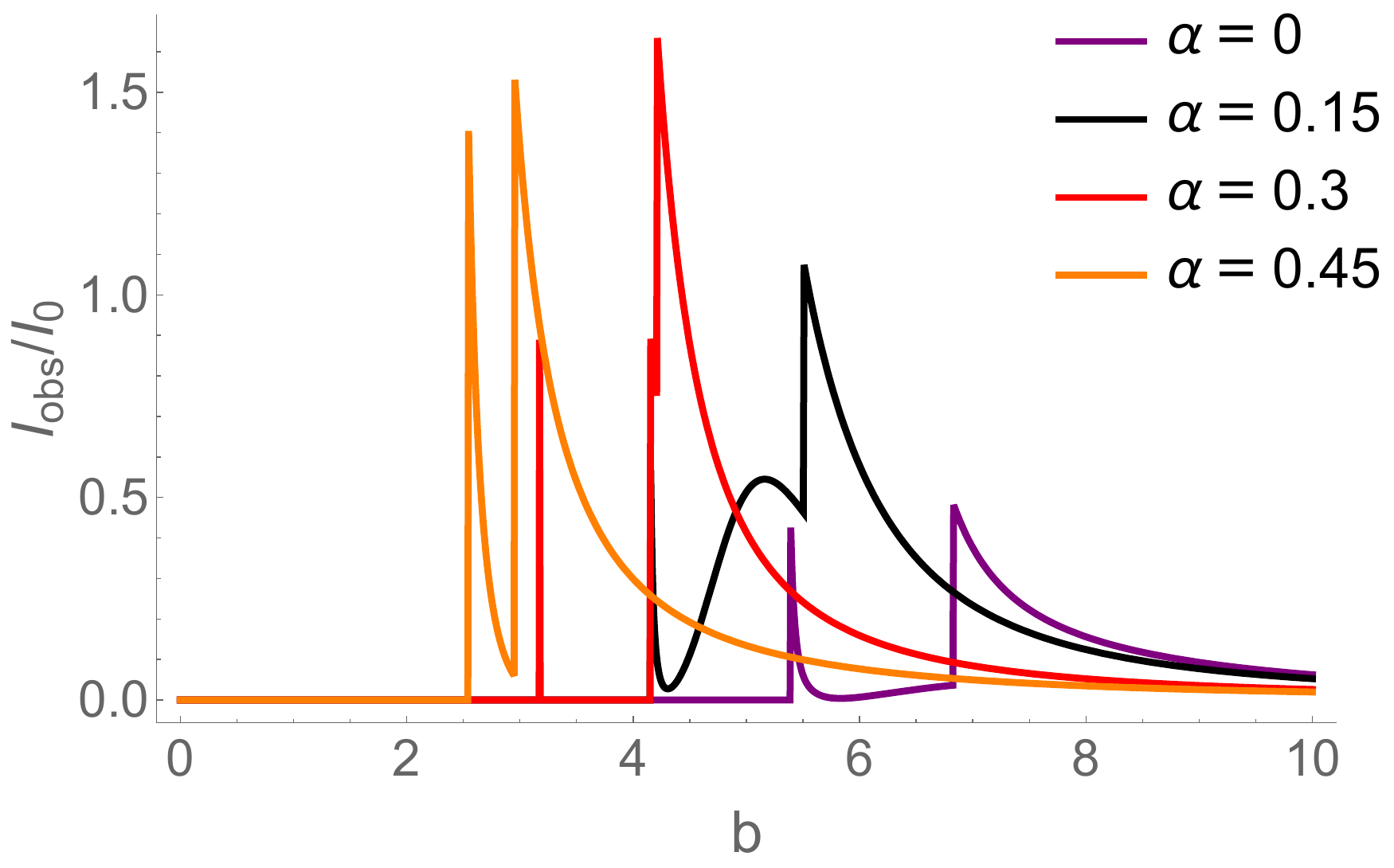}
\end{minipage}
\begin{minipage}{0.3\textwidth}
    \centering
    \begin{minipage}{0.75\linewidth}
        \includegraphics[width=\linewidth]{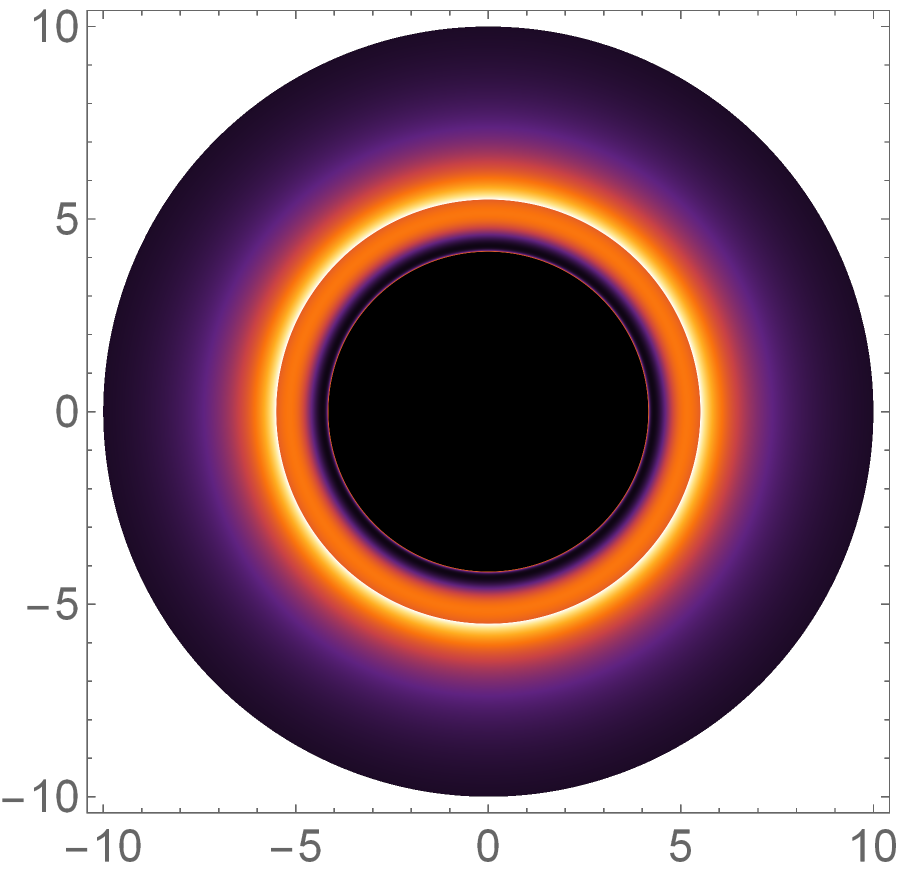} 
    \end{minipage}
    \begin{minipage}{0.14\linewidth}
        \includegraphics[width=\linewidth]{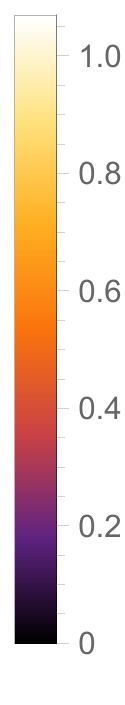} 
    \end{minipage}
\end{minipage}
\begin{minipage}{0.34\textwidth}
    \includegraphics[width=\linewidth]{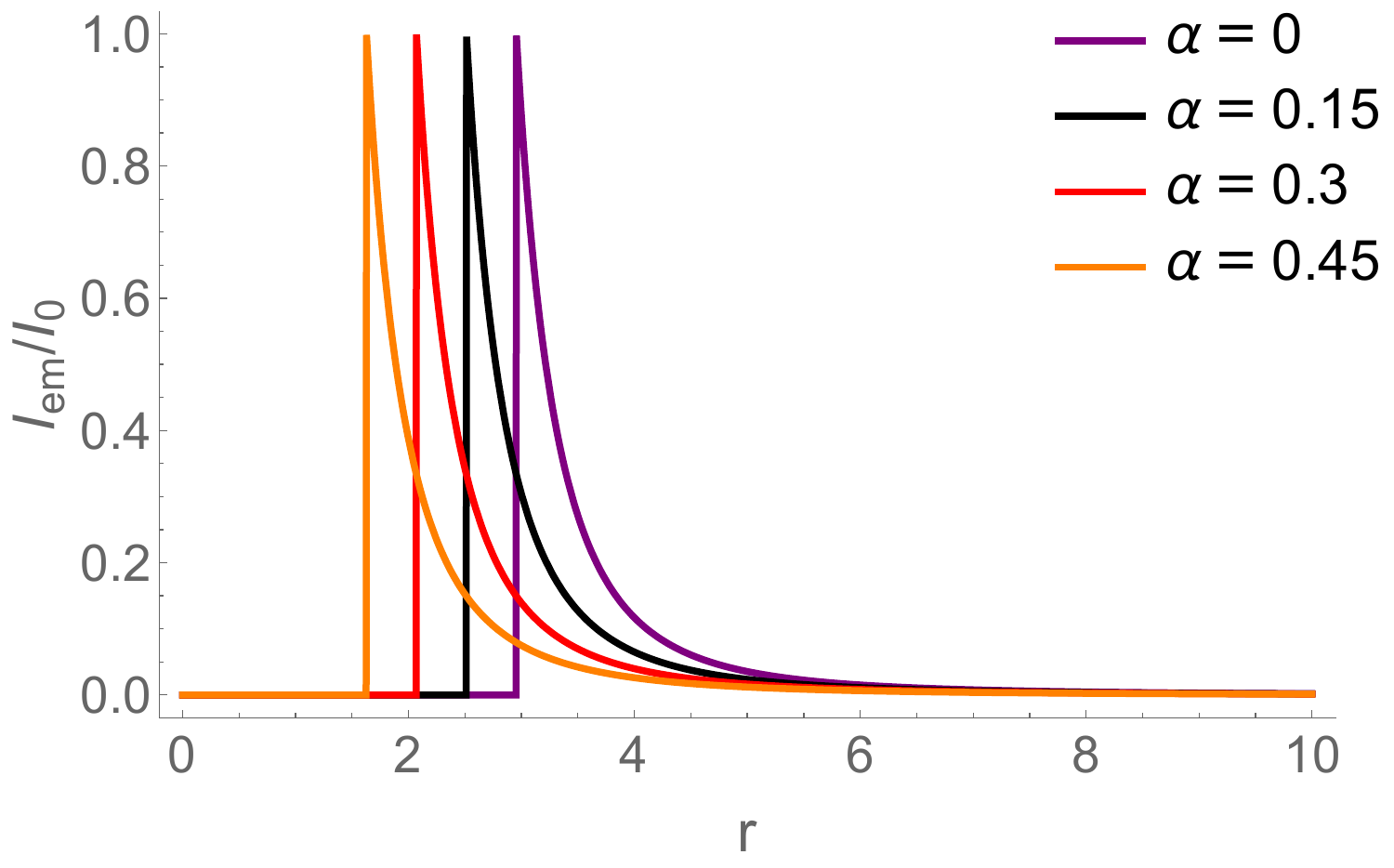}
\end{minipage}
\begin{minipage}{0.34\textwidth}
    \includegraphics[width=\linewidth]{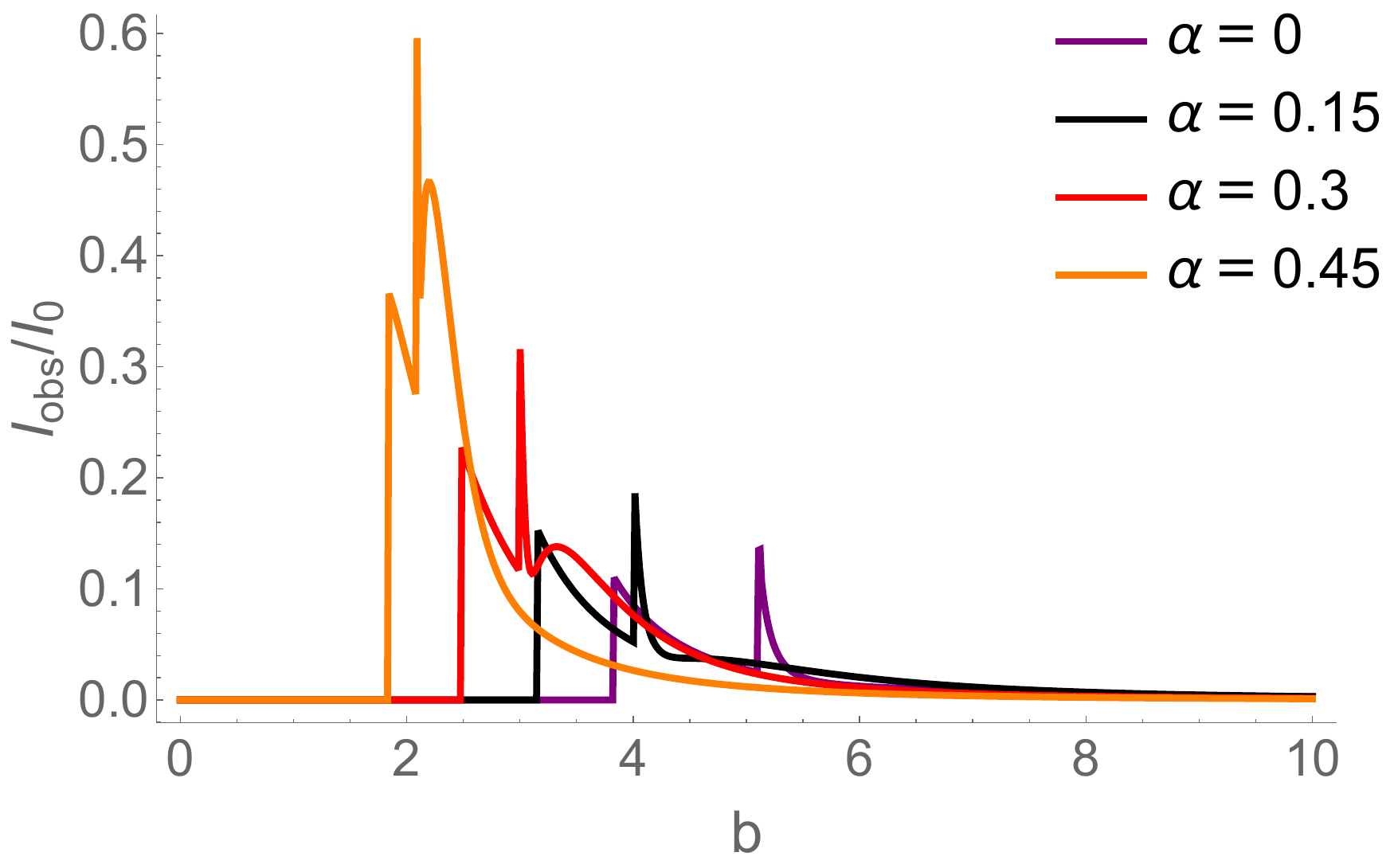}
\end{minipage}
\begin{minipage}{0.3\textwidth}
    \centering
    \begin{minipage}{0.75\linewidth}
        \includegraphics[width=\linewidth]{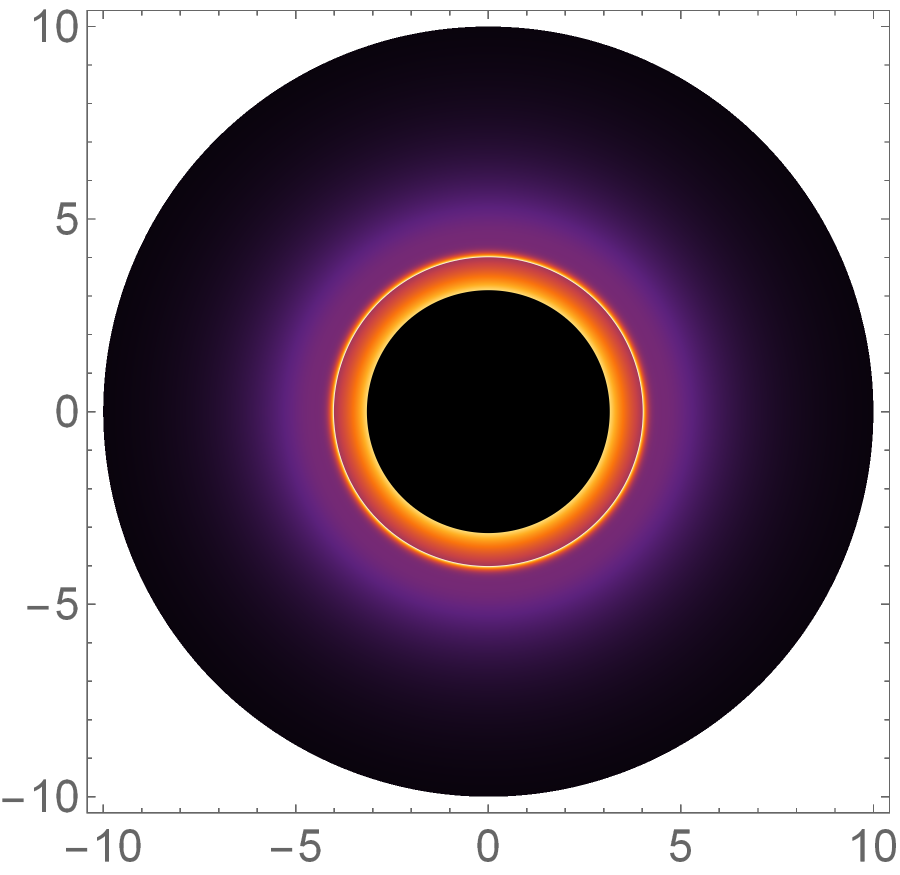} 
    \end{minipage}
    \begin{minipage}{0.18\linewidth}
        \includegraphics[width=\linewidth]{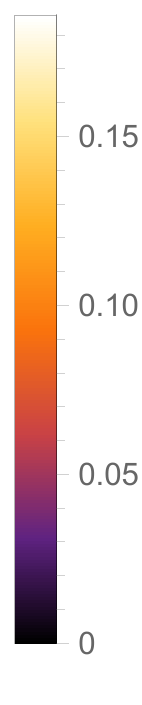} 
    \end{minipage}
\end{minipage}

\begin{minipage}{0.34\textwidth}
    \includegraphics[width=\linewidth]{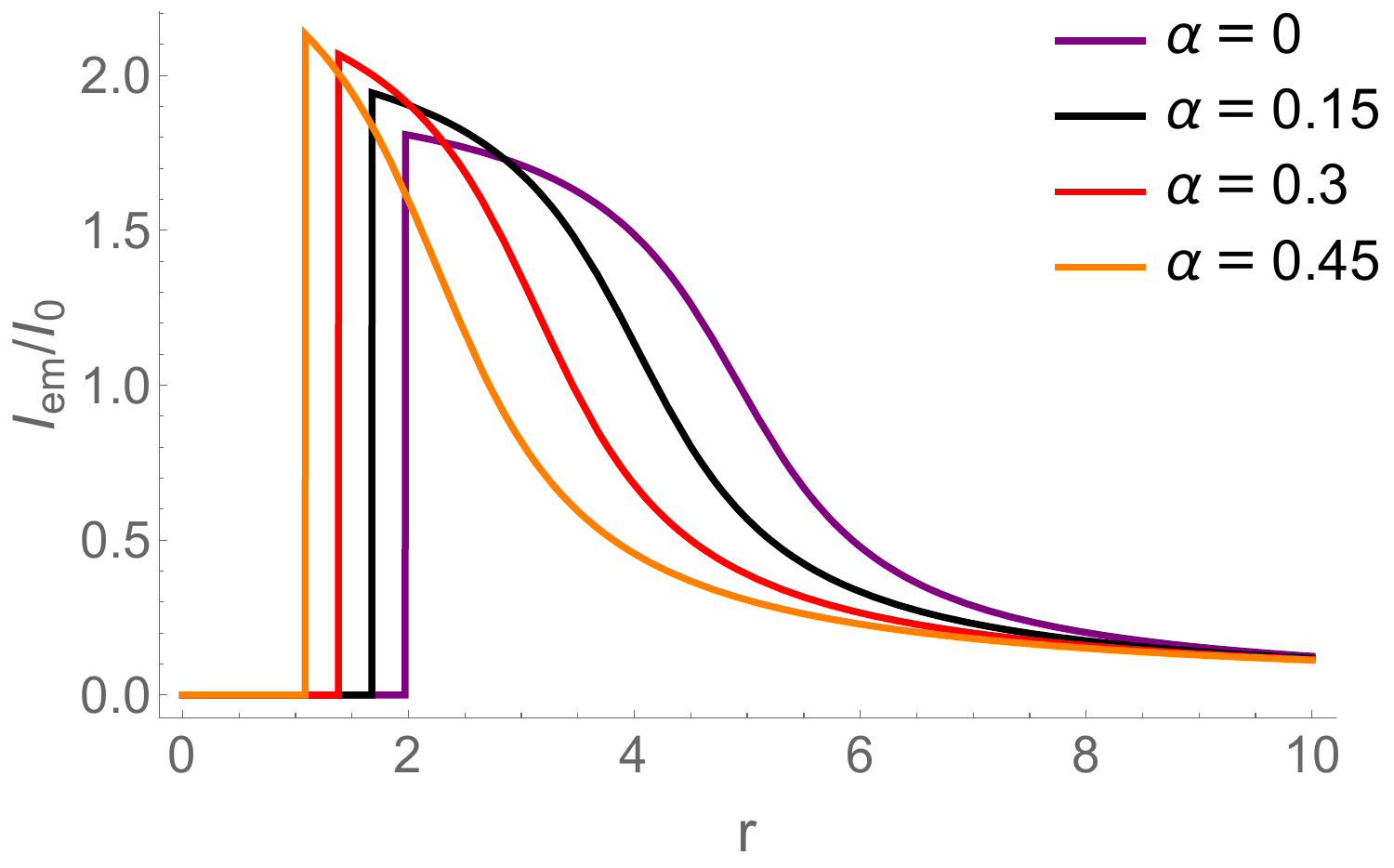}
\end{minipage}
\begin{minipage}{0.34\textwidth}
    \includegraphics[width=\linewidth]{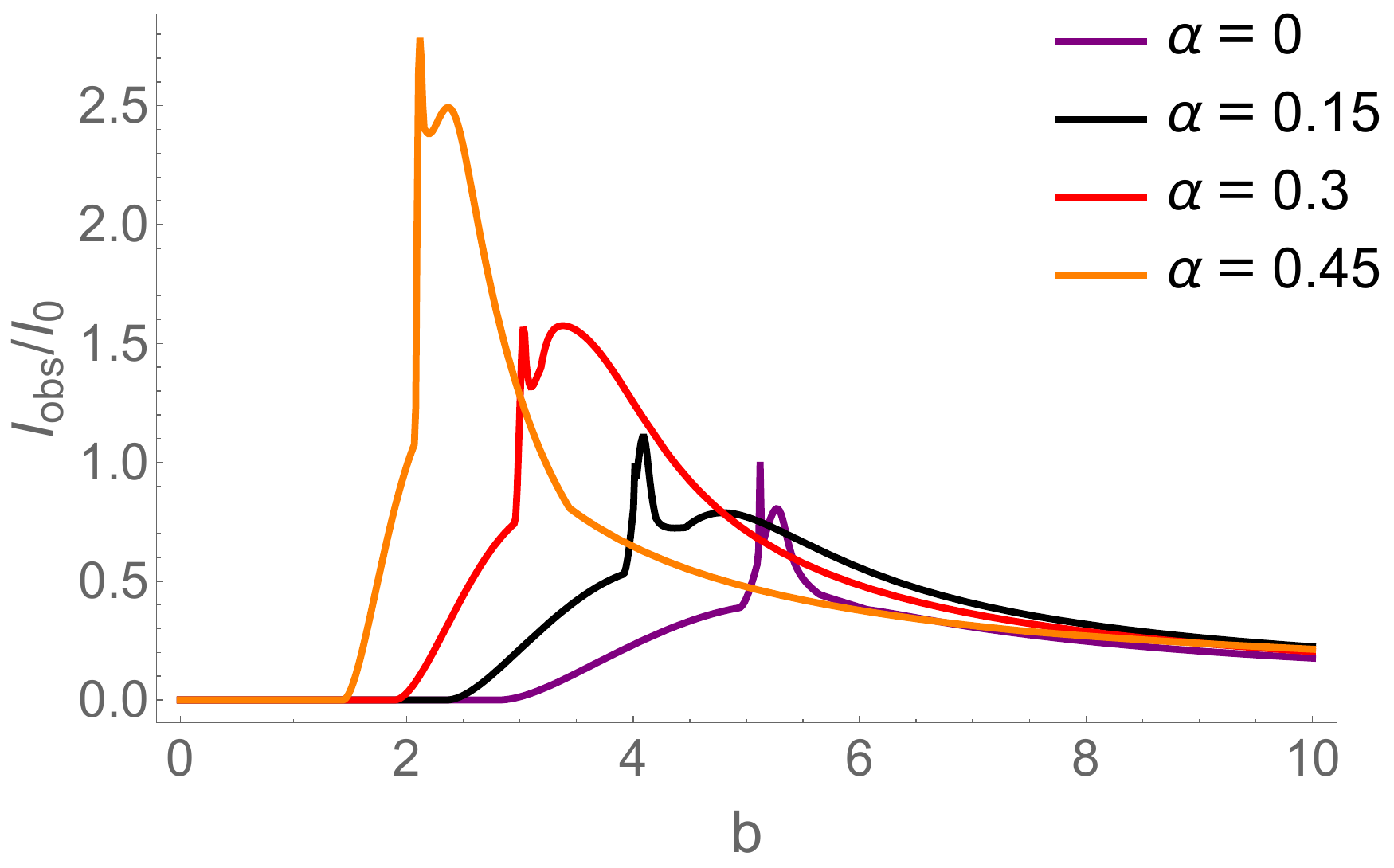}
\end{minipage}
\begin{minipage}{0.3\textwidth}
    \centering
    \begin{minipage}{0.75\linewidth}
        \includegraphics[width=\linewidth]{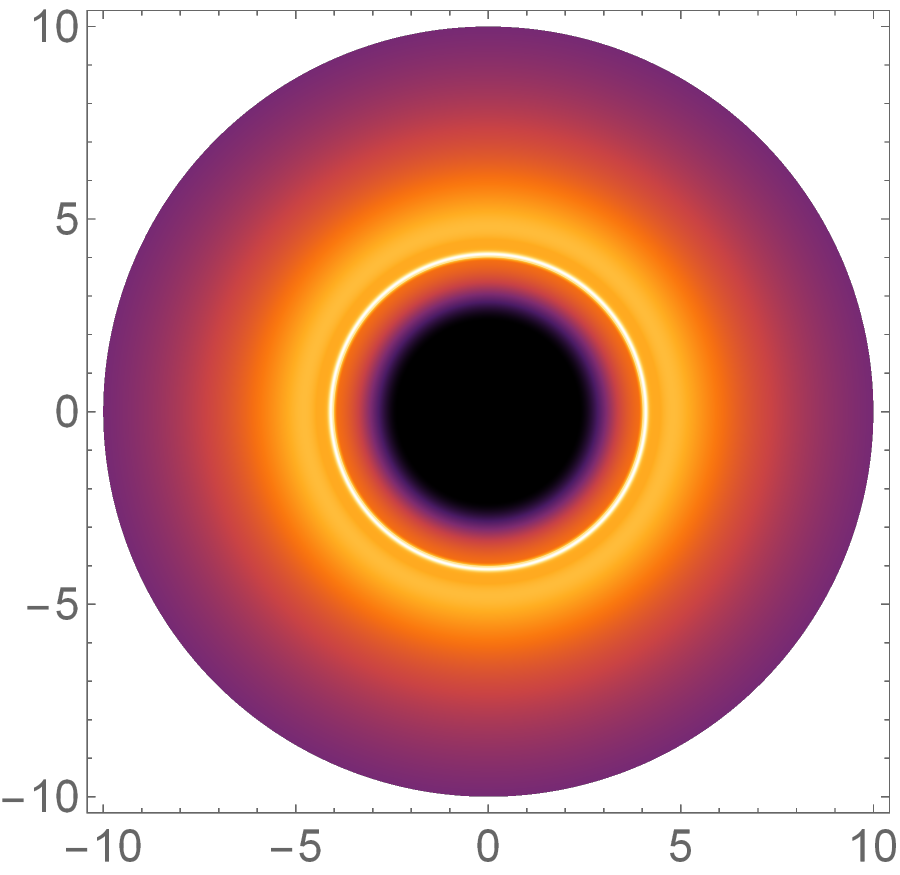} 
    \end{minipage}
    \begin{minipage}{0.14\linewidth}
        \includegraphics[width=\linewidth]{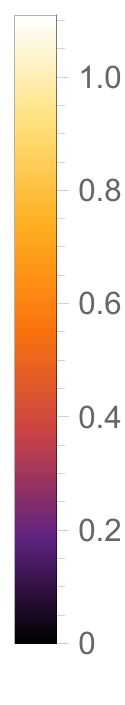} 
    \end{minipage}
\end{minipage}

\caption{Emission intensity (first column), observed intensity (second column), and density maps (third column, with $\alpha = 0.15$) of the thin accretion disk around the black hole for different values of the Lorentz parameter $\alpha$, with $\lambda = 0.005$ fixed.}
\centering
\label{fig:4}
\end{figure*}

As shown in Fig.~\ref{fig:3}, the transfer functions are plotted, where the black curve represents the first transfer function with the impact parameter range $b \in (b_1^{-},\, \infty)$, the yellow segment represents the second transfer function with $b \in (b_2^{-},\, b_2^{+})$, and the red curve represents the third transfer function with $b \in (b_3^{-},\, b_3^{+})$. It can be seen that as the order $i$ of the transfer function increases, the allowed range of $b$ becomes progressively narrower, and its contribution to the observed intensity correspondingly decreases. For higher-order transfer functions ($i \geq 4$), the contribution can be safely neglected.

We discuss the following three accretion disk emission models and their corresponding images.

\begin{equation}
I_{\rm em}(r) := 
\begin{cases}
I_0 \left[ \frac{1}{r - (r_{\text{isco}} - 1)} \right]^2, & r > r_{\text{isco}}, \\
0, & r \le r_{\text{isco}}.
\end{cases}
\label{equ:19}
\end{equation}

\begin{equation}
I_{\rm em}(r) := 
\begin{cases}
I_0 \left[ \frac{1}{r - (r_{\text{ph}} - 1)} \right]^3, & r > r_{\text{ph}}, \\
0, & r \le r_{\text{ph}}.
\end{cases}
\label{equ:20}
\end{equation}

\begin{equation}
I_{\rm em}(r) := 
\begin{cases}
I_0 \frac{\frac{\pi}{2} - \arctan\bigl[r - (r_{\text{isco}} - 1)\bigr]}{\frac{\pi}{2} - \arctan\bigl[r_h - (r_{\text{isco}} - 1)\bigr]}, & r > r_h, \\
0, & r \le r_h.
\label{equ:21}
\end{cases}
\end{equation}

\begin{figure*}[t]
\centering
\begin{minipage}{0.34\textwidth}
    \includegraphics[width=\linewidth]{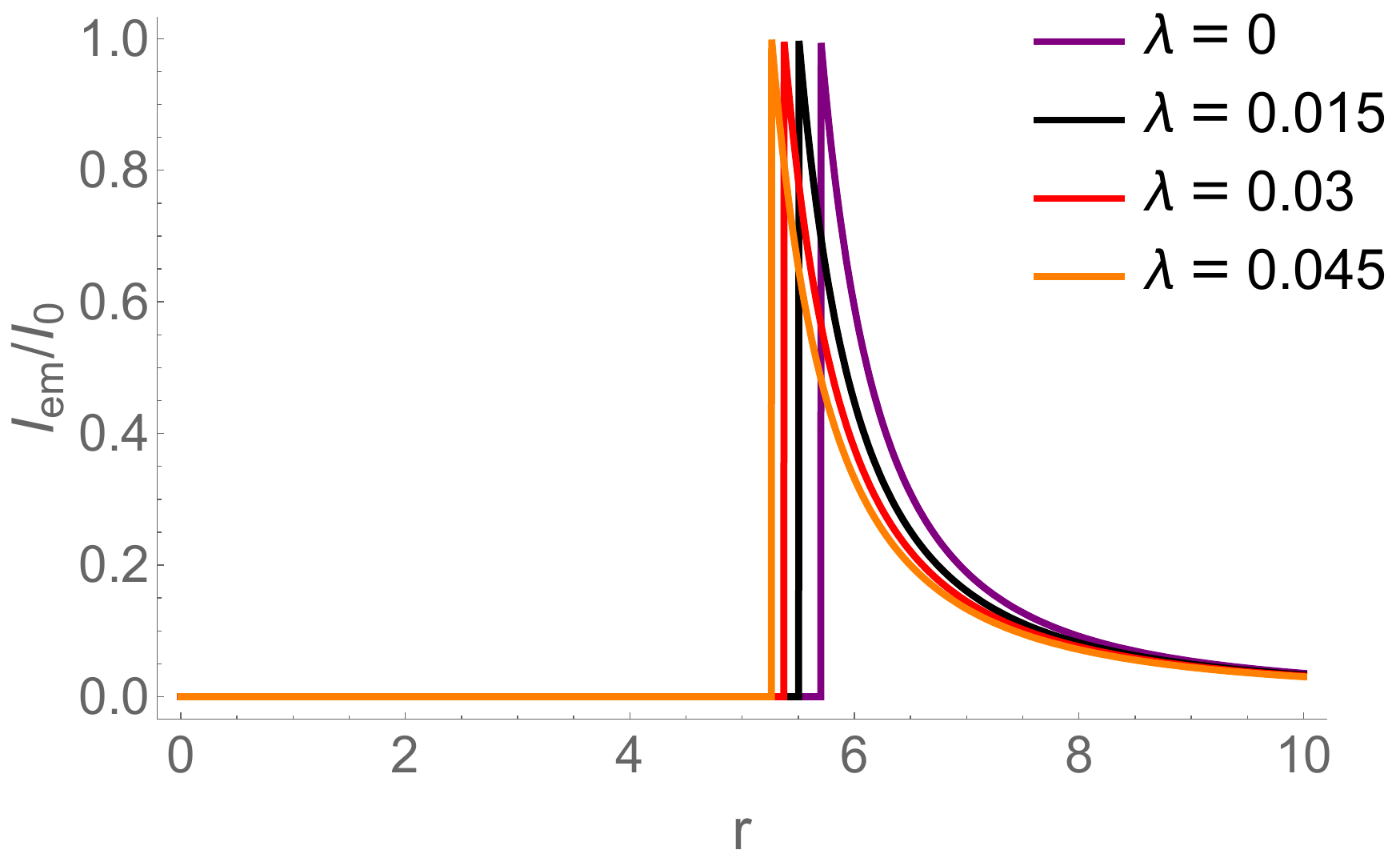}
\end{minipage}
\begin{minipage}{0.34\textwidth}
    \includegraphics[width=\linewidth]{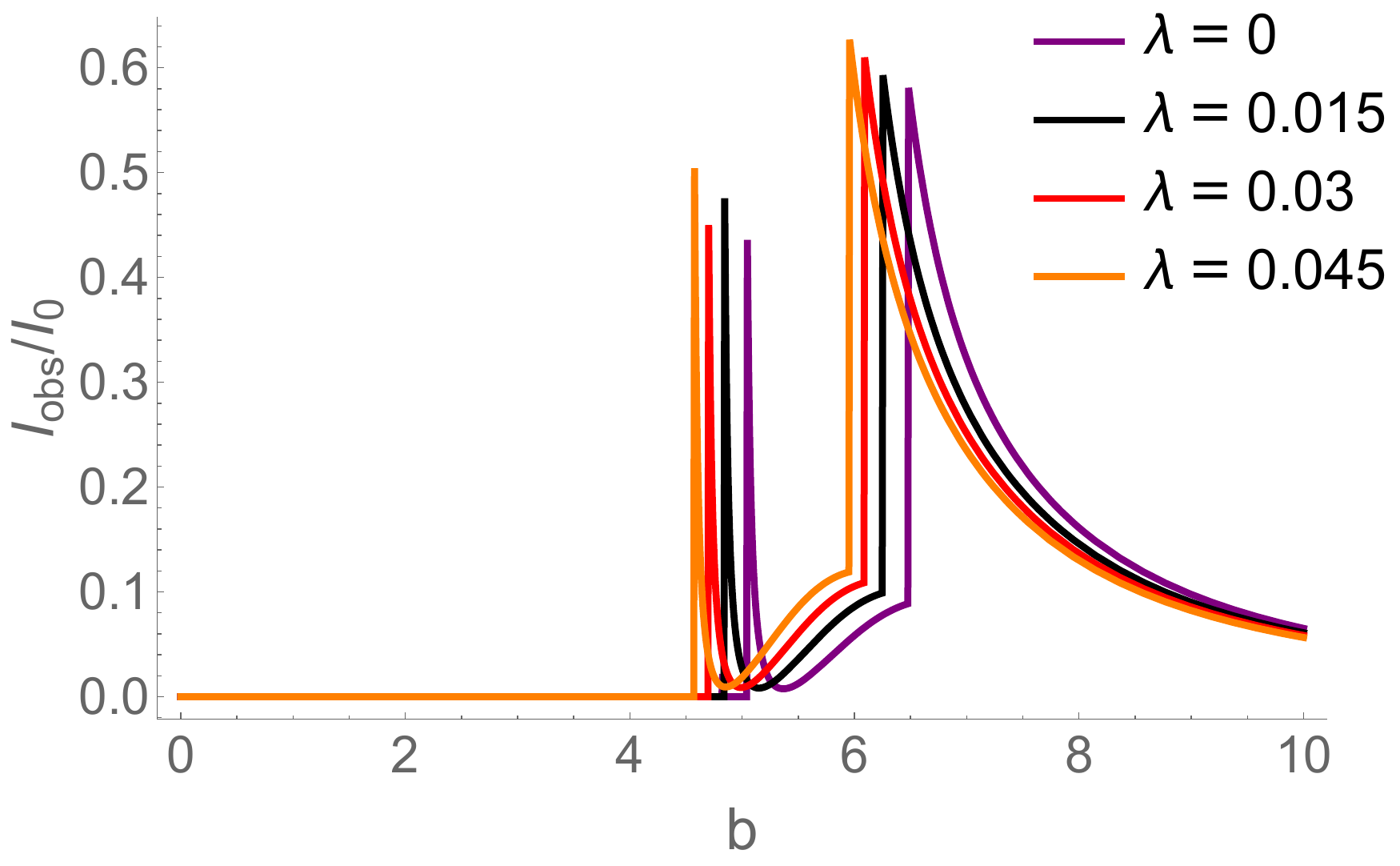}
\end{minipage}
\begin{minipage}{0.3\textwidth}
    \centering
    \begin{minipage}{0.75\linewidth}
        \includegraphics[width=\linewidth]{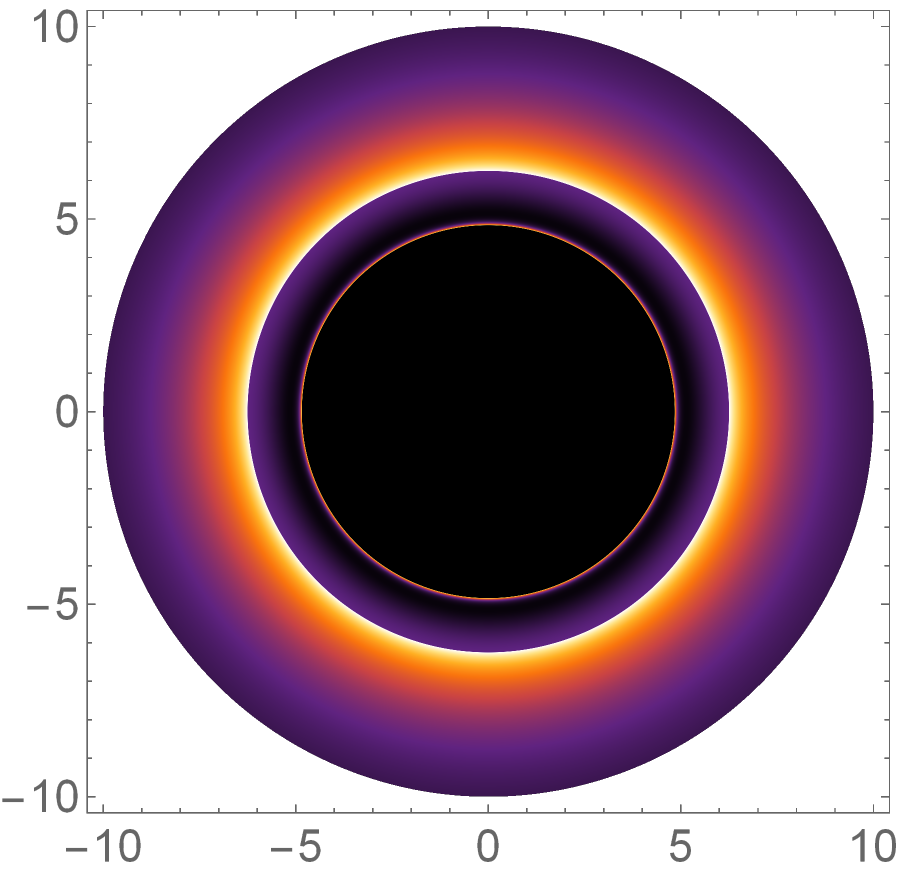} 
    \end{minipage}
    \begin{minipage}{0.14\linewidth}
        \includegraphics[width=\linewidth]{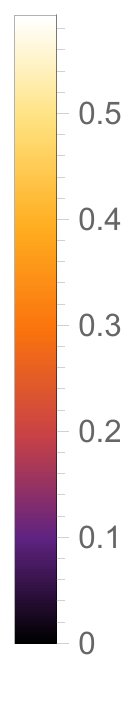} 
    \end{minipage}
\end{minipage}
\begin{minipage}{0.34\textwidth}
    \includegraphics[width=\linewidth]{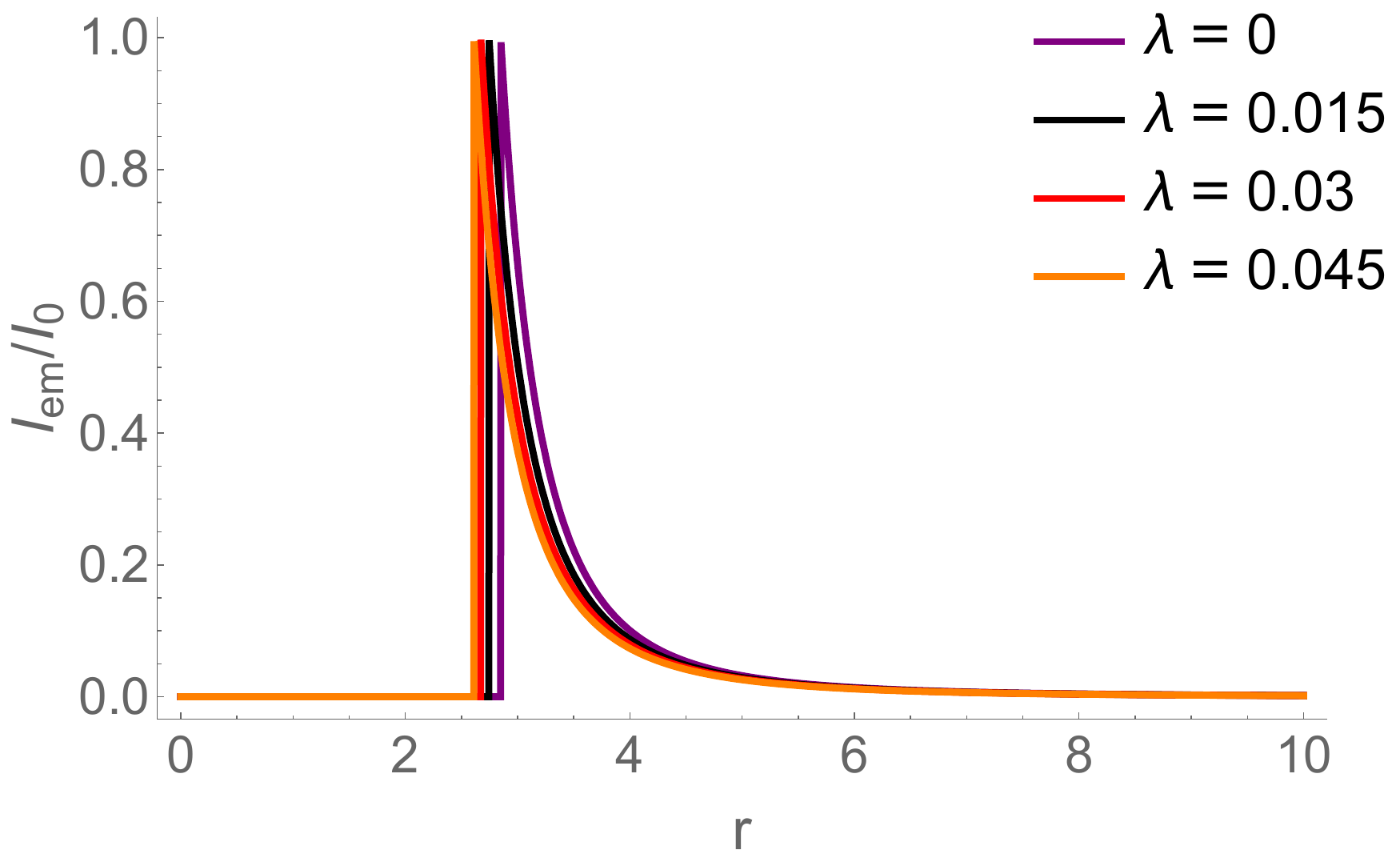}
\end{minipage}
\begin{minipage}{0.34\textwidth}
    \includegraphics[width=\linewidth]{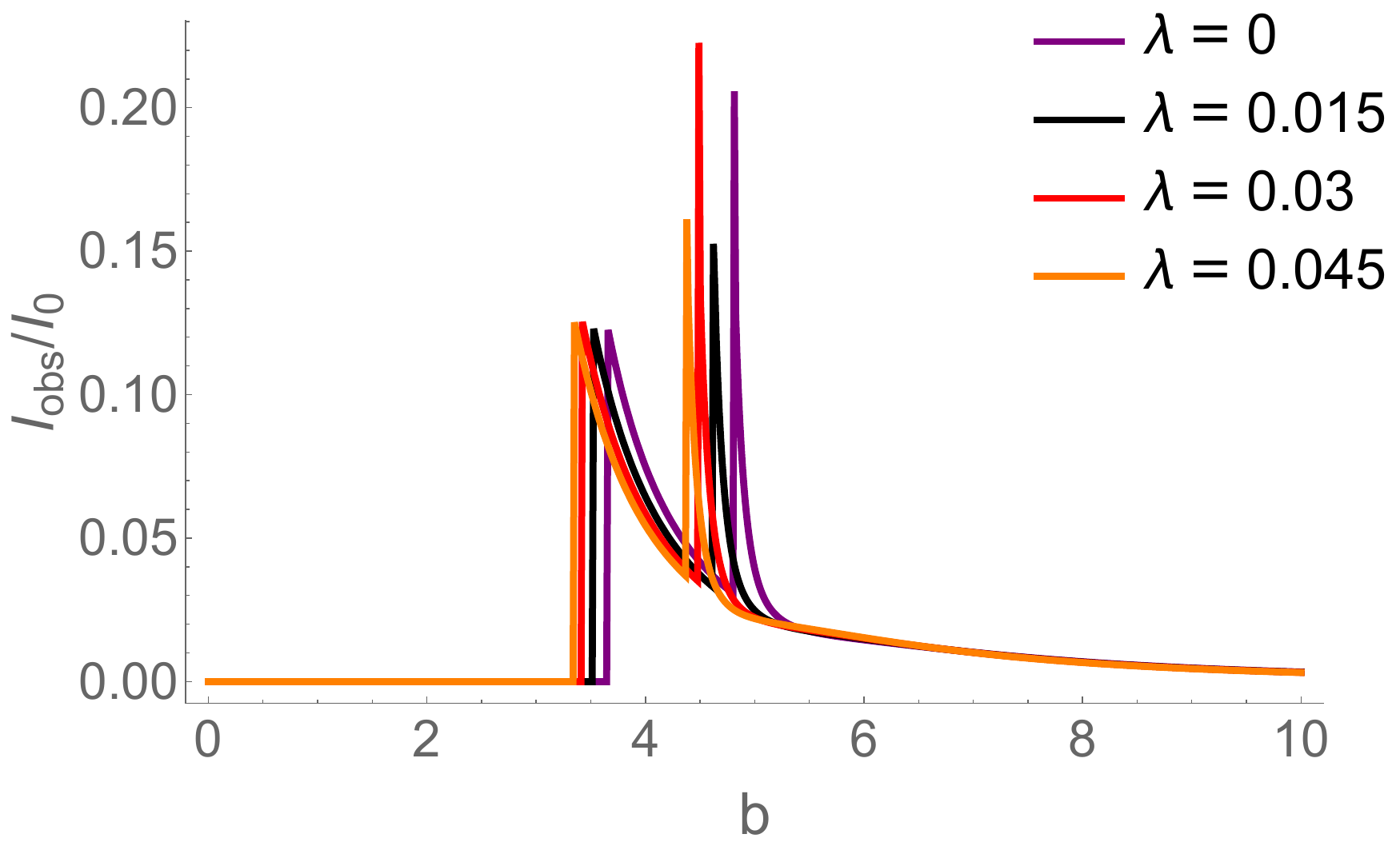}
\end{minipage}
\begin{minipage}{0.3\textwidth}
    \centering
    \begin{minipage}{0.75\linewidth}
        \includegraphics[width=\linewidth]{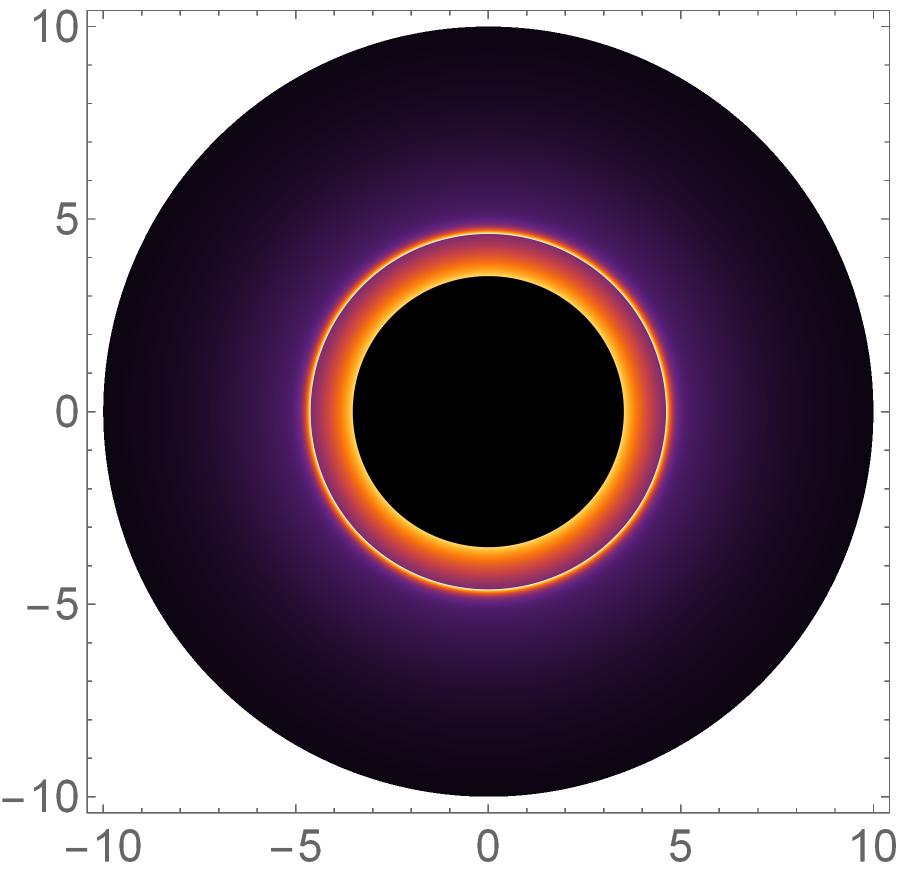} 
    \end{minipage}
    \begin{minipage}{0.18\linewidth}
        \includegraphics[width=\linewidth]{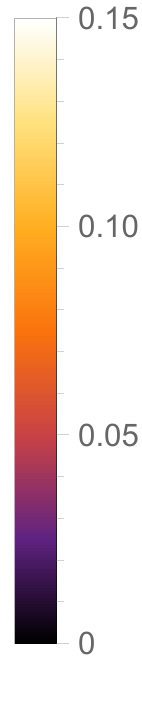} 
    \end{minipage}
\end{minipage}

\begin{minipage}{0.34\textwidth}
    \includegraphics[width=\linewidth]{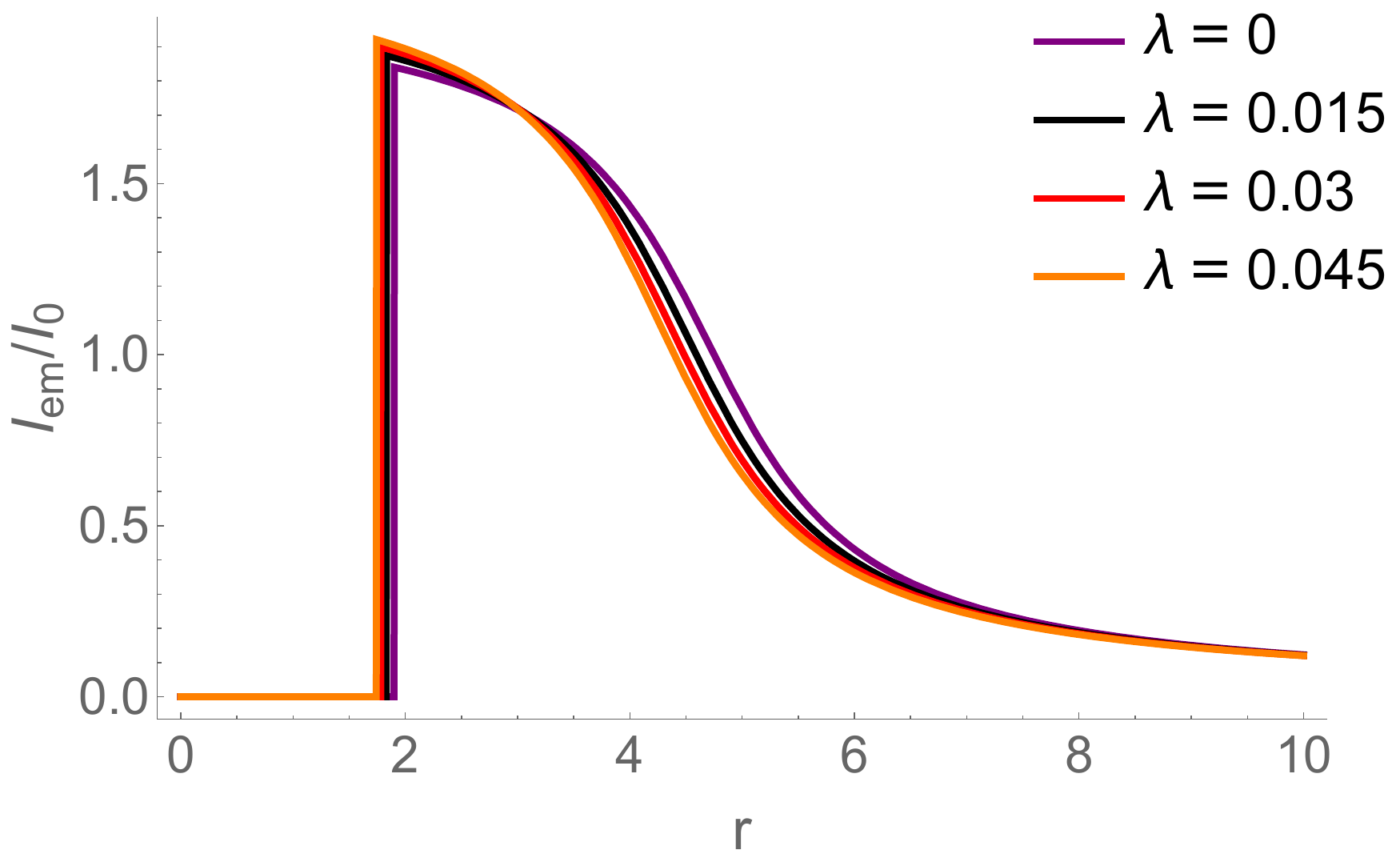}
\end{minipage}
\begin{minipage}{0.34\textwidth}
    \includegraphics[width=\linewidth]{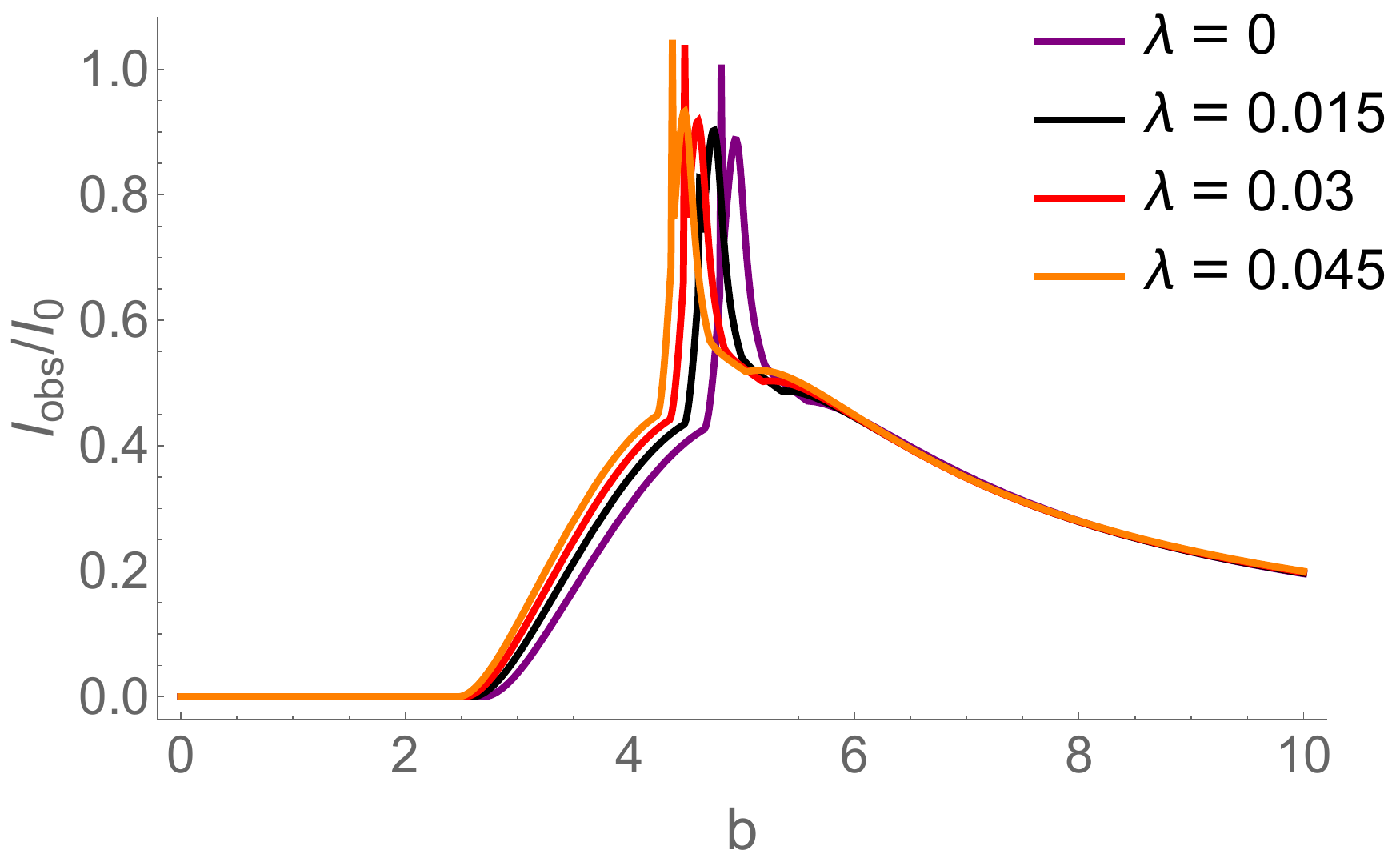}
\end{minipage}
\begin{minipage}{0.3\textwidth}
    \centering
    \begin{minipage}{0.75\linewidth}
        \includegraphics[width=\linewidth]{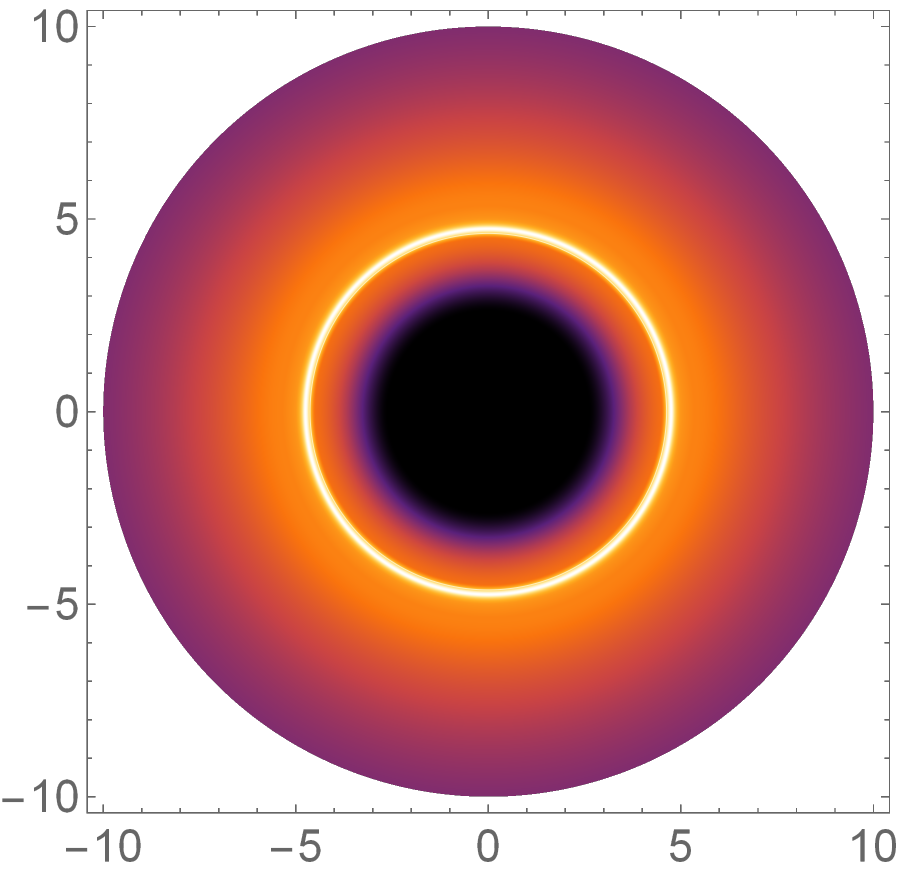} 
    \end{minipage}
    \begin{minipage}{0.14\linewidth}
        \includegraphics[width=\linewidth]{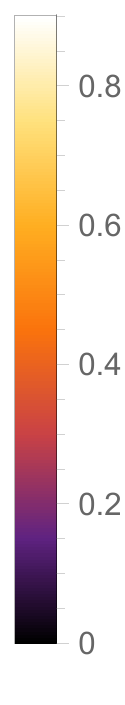} 
    \end{minipage}
\end{minipage}

\caption{Emission intensity (first column), observed intensity (second column), and density maps (third column, with $\lambda = 0.015$) of the thin accretion disk around the black hole for different values of the dark matter parameter $\lambda$, with $\alpha = 0.05$ fixed.}
\centering
\label{fig:5}
\end{figure*}

As shown in Fig.~\ref{fig:4} and Fig.~\ref{fig:5}, the thin accretion disk structures of the black hole are presented for different values of the Lorentz parameter $\alpha$ and the dark matter parameter $\lambda$. The first column represents the emission intensity, the second column shows the observed intensity, and the third column displays the corresponding density maps of the observed intensity. The emission intensity of these models reaches its maximum at $r_{\mathrm{isco}}$ (first row), $r_{\mathrm{ph}}$ (second row), and $r_{\rm h}$ (third row), respectively. It vanishes inside these radii, while outside these regions it gradually decreases with increasing radius.

All emission images exhibit a single-peak structure, whereas the corresponding observed images display multi-peak features, typically showing a double-peak structure. In fact, from a theoretical point of view, the observed image should contain three peaks. From right to left, these correspond to the first-order transfer function (direct ring), the second-order transfer function (lensing ring), and the third-order transfer function (photon ring), respectively. In Fig.~\ref{fig:4} and Fig.~\ref{fig:5}, for the first two emission models, the observed images exhibit a clear double-peak structure due to the overlap between the lensing ring and the photon ring. For the third model, all three peaks contribute in the vicinity of $b_c$.

Furthermore, as $\alpha$ and $\lambda$ increase, both the emission and observed intensity profiles shift toward smaller values of the impact parameter $b$. These differences and trends in the observed intensity may provide useful insights for interpreting observational features of different accretion disks.

\section{Ringdown}
\label{sec:qnm}

\subsection{Perturbation Equations and Effective Potentials}
While the photon-ring structures reveal the behavior of light propagation near the black hole, they do not capture the dynamical response of the spacetime. When subjected to perturbations, the black hole relaxes toward equilibrium through the emission of damped gravitational waves during the ringdown phase. The characteristics of this signal depend on the underlying geometry and thus provide an independent probe of the system. In the following, we investigate the ringdown properties of black holes in a Kalb--Ramond field coupled to perfect fluid dark matter.

When a black hole is subjected to external perturbations, its dynamical response can be characterized by studying the propagation of test fields with different spins in the background spacetime. According to the spin of the field, the typical types of perturbations include scalar fields ($s=0$), electromagnetic fields ($s=1$), and gravitational fields ($s=2$). The scalar field perturbation is governed by the Klein--Gordon equation in curved spacetime,

\begin{equation}
 \frac{1}{\sqrt{-g}} \partial_{\mu} (\sqrt{-g} g^{\mu\nu} \partial_{\nu} \Phi) = 0
\label{equ:22}
\end{equation}
Electromagnetic perturbations satisfy the source-free Maxwell equations,

\begin{equation}
\frac{1}{\sqrt{-g}} \partial_{\nu} \left( F_{\rho \sigma} g^{\rho \mu} g^{\sigma \nu} \sqrt{-g} \right) = 0
\label{equ:23}
\end{equation}
where $F_{\mu\nu}$ is the electromagnetic field tensor. In contrast, gravitational perturbations correspond to small deviations of the metric tensor itself, and their dynamics are determined by the linearized Einstein equations.

\begin{figure*}[t!]
    \begin{minipage}[b]{0.32\textwidth}
        \centering
        \includegraphics[width=\textwidth]{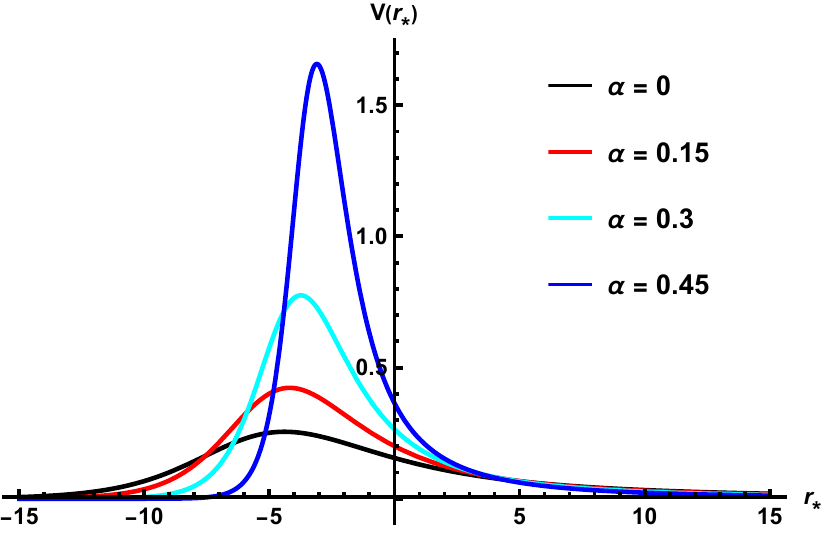}
    \end{minipage}
    \begin{minipage}[b]{0.32\textwidth}
        \centering
        \includegraphics[width=\textwidth]{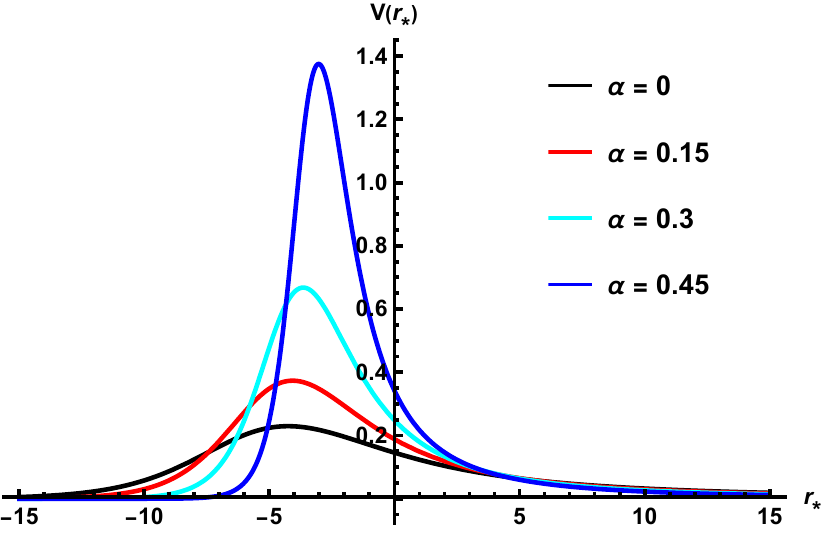}
    \end{minipage}%
    \begin{minipage}[b]{0.32\textwidth}
        \centering
        \includegraphics[width=\textwidth]{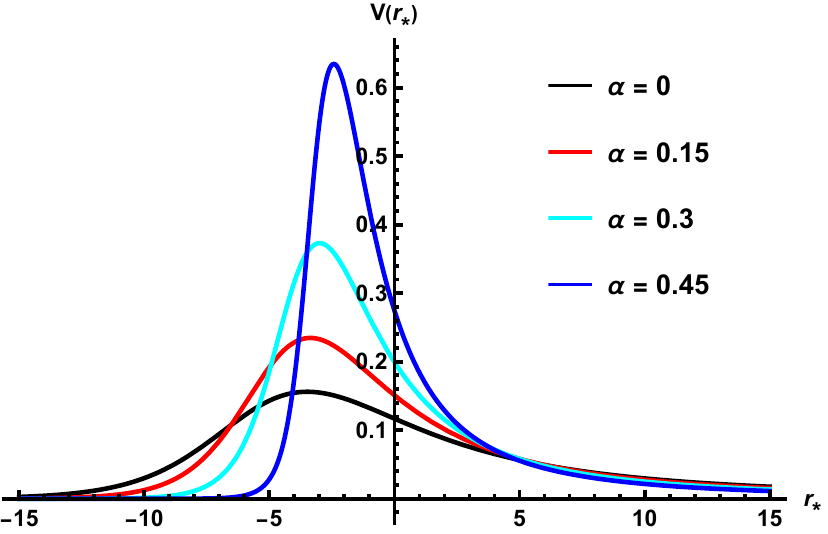}
    \end{minipage}
    \caption{Effective potentials for different perturbations. From left to right: scalar field, electromagnetic field, and axial gravitational field, with $\lambda=0.005, l=2$.}
    \label{fig:6}
\end{figure*}

\begin{figure*}[t!]
    \begin{minipage}[b]{0.32\textwidth}
        \centering
        \includegraphics[width=\textwidth]{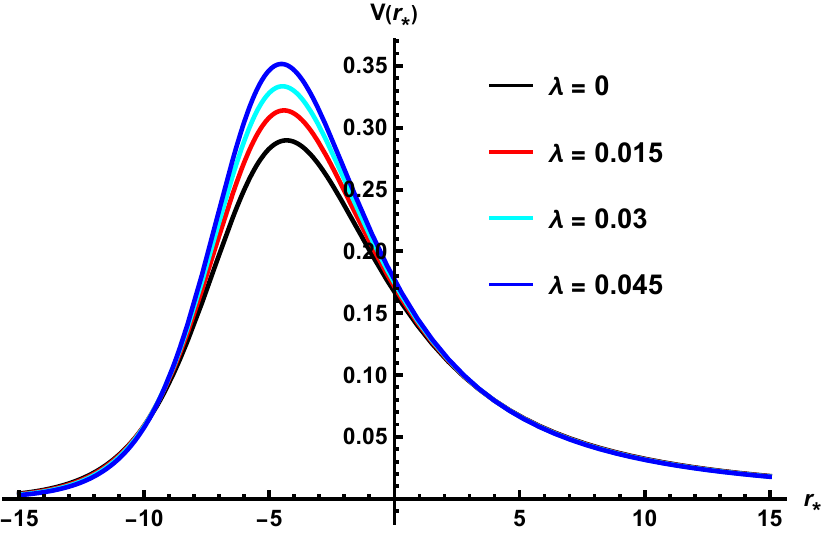}
    \end{minipage}
    \begin{minipage}[b]{0.32\textwidth}
        \centering
        \includegraphics[width=\textwidth]{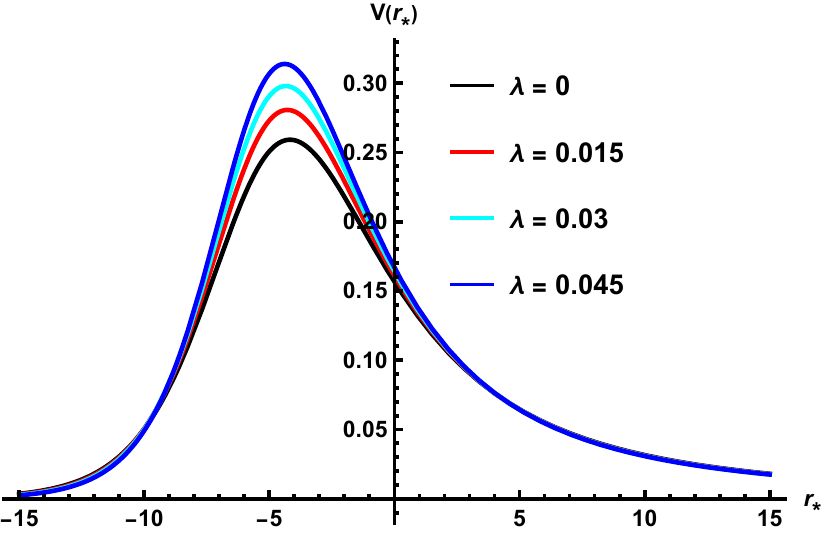}
    \end{minipage}%
    \begin{minipage}[b]{0.32\textwidth}
        \centering
        \includegraphics[width=\textwidth]{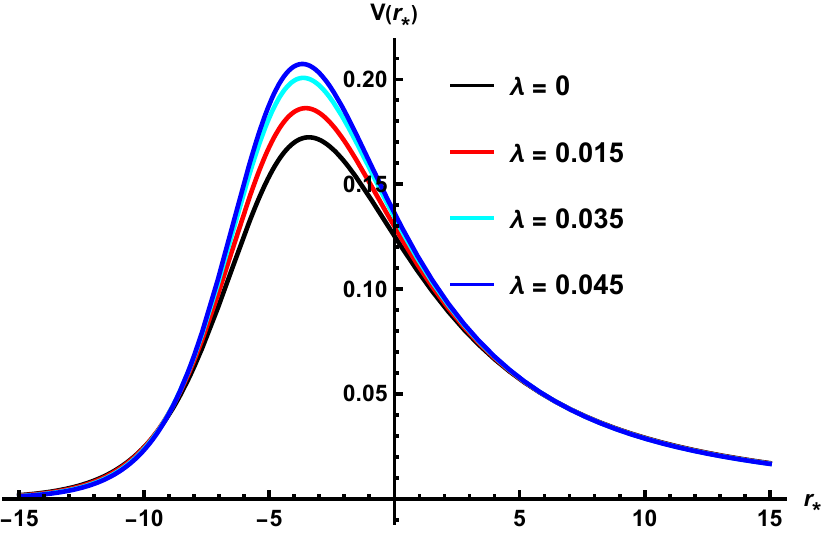}
    \end{minipage}
    \caption{Effective potentials for different perturbations. From left to right: scalar field, electromagnetic field, and axial gravitational field, with $\alpha = 0.05, l=2$ .}
    \label{fig:7}
\end{figure*}

Gravitational perturbations can be further decomposed according to parity into axial (odd-parity) and polar (even-parity) modes~\cite{Kokkotas:1999bd}. Since the axial perturbations possess a relatively simpler mathematical structure and can typically be reduced to a single wave equation, we focus on the axial gravitational perturbations in this work.For gravitational perturbations, the spacetime can be decomposed into a background part and a small perturbative part. Within the framework of linear perturbation theory, the metric can be written as:

\begin{equation}
g_{\mu\nu}=\bar{g}_{\mu\nu}+h_{\mu\nu}
\label{equ:24}
\end{equation}

In a spherically symmetric background, the perturbation field can be written as $\Phi(t,r,\theta,\phi)=Y(\theta,\phi)\,\frac{\psi(t,r)}{r}$, and both types of perturbations can be reduced to a one-dimensional radial wave equation through separation of variables .

\begin{equation}
\frac{d^2\varphi(r_*)}{dr_*^2}
+
\left[\omega^2 - V(r_*)\right]\varphi(r_*)=0
\label{equ:25}
\end{equation}

$r_*$ is the tortoise coordinate, whose relation to the radial coordinate $r$ is given by $dr_*=\frac{dr}{f(r)}$; the radial function can be written as $\psi(t,r_*)=e^{-i\omega t}\,\varphi(r_*)$. The effective potential in Eq.~\ref{equ:25} is given by:

\begin{equation}
V(r_*) \equiv V\!\big(r(r_*)\big)
=
f(r)\left[
\frac{l(l+1)}{r^2}
+
\frac{1-s^2}{r}\,\frac{df(r)}{dr}
\right]
\label{equ:26}
\end{equation}

In Eq.~\ref{equ:26}, $l$ is the multipole quantum number, and $s$ denotes the spin of the perturbation field. When $s=0, 1,$ and $2$, they correspond to scalar, electromagnetic, and gravitational perturbations, respectively. For the black hole considered in this work under these three types of perturbations, substituting $f(r) = \frac{1}{1 - \alpha} - \frac{2M}{r} + \frac{\lambda}{r} \log \frac{r}{|\lambda|}$ into the above expression yields the corresponding effective potentials as follows: 

\begin{equation}
\begin{aligned}
V(r)|_{s = 0}&=\left(\frac{1}{1-\alpha}-\frac{2M}{r}
+\frac{\lambda \log\!\left(\frac{r}{|\lambda|}\right)}{r}\right)\\
&\left[
\frac{l(1+l)}{r^2}
+
\frac{1}{r}
\left(
\frac{2M}{r^2}
+\frac{\lambda}{r^2}
-\frac{\lambda \log\!\left(\frac{r}{|\lambda|}\right)}{r^2}
\right)
\right]
\end{aligned}
\label{equ:27}
\end{equation}

\begin{equation}
V(r)|_{s = 1}=\left(\frac{1}{1-\alpha}-\frac{2M}{r}
+\frac{\lambda \log\!\left(\frac{r}{|\lambda|}\right)}{r}\right)
\left[\frac{l(1+l)}{r^2}\right]
\label{equ:28}
\end{equation}

\begin{equation}
\begin{aligned}
V(r)|_{s =2}&=\left(\frac{1}{1-\alpha}-\frac{2M}{r}
+\frac{\lambda \log\!\left(\frac{r}{|\lambda|}\right)}{r}\right)\\&
\left[
\frac{l(1+l)}{r^2}
-
\frac{3}{r}
\left(
\frac{2M}{r^2}
+\frac{\lambda}{r^2}
-\frac{\lambda \log\!\left(\frac{r}{|\lambda|}\right)}{r^2}
\right)
\right]
\end{aligned}
\label{equ:29}
\end{equation}

As shown in Fig.~\ref{fig:6} and Fig.~\ref{fig:7}, the effective potentials of a black hole under three types of field perturbations are presented for different values of the Lorentz parameter $\alpha$ and the dark matter parameter $\lambda$. It can be seen that perturbation fields with different spins respond differently to the background spacetime. Although the effective potentials for scalar, electromagnetic, and gravitational perturbations share a similar overall profile, their peak heights and steepness are significantly different. Under the same parameter conditions, the scalar field exhibits the highest potential barrier, followed by the electromagnetic field, while the gravitational perturbation corresponds to the lowest barrier.

Moreover, in all cases the effective potentials display a typical single-peak barrier structure. As either $\alpha$ or $\lambda$ increases, the height of the barrier increases accordingly. In general, a higher barrier provides stronger confinement for the perturbation waves, shortening the characteristic time for repeated reflections within the potential well and thus leading to a higher oscillation frequency. Therefore, it can be expected that the real part of the quasinormal mode frequency increases with increasing $\alpha$ or $\lambda$. In addition, a wider barrier implies a lower tunneling probability for the perturbation waves, resulting in slower energy dissipation and a longer-lived ringdown signal, which corresponds to a smaller absolute value of the imaginary part of the quasinormal mode frequency. From the figures, it can be observed that the barrier width decreases as $\alpha$ increases, while the influence of $\lambda$ on the barrier width is relatively weak.

In summary, the qualitative behavior of the quasinormal modes with respect to the model parameters can be inferred from the shape of the effective potentials, whereas the precise frequencies and damping rates must be determined through numerical calculations.

\subsection{Numerical methods and results}

In the previous section, we have shown that both matter-field perturbations and gravitational perturbations can be reduced, via separation of variables, to the one-dimensional radial wave Eq.~\ref{equ:25}. In the asymptotic regions $r_* \to \pm \infty$, the effective potential vanishes and Eq.~\ref{equ:25} reduces to plane-wave solutions. The quasinormal mode frequency $\omega$ is a discrete complex eigenvalue determined by specific boundary conditions, which require that the perturbation be purely ingoing at the event horizon and purely outgoing at spatial infinity:
\begin{equation}
\varphi(r_*) \sim
\begin{cases}
e^{-i\omega r_*}, & r_* \to -\infty, \\
e^{+i\omega r_*}, & r_* \to +\infty .
\end{cases}
\label{equ:30}
\end{equation}
Since exact analytical solutions are generally unavailable, the quasinormal mode frequencies are typically obtained using numerical or semi-analytical methods. In this work, we employ two complementary approaches to investigate the oscillatory properties of the system: the higher-order WKB semi-analytical method and the time-domain evolution method.

\begin{figure*}[t!]
    \begin{minipage}[b]{0.32\textwidth}
        \centering
        \includegraphics[width=\textwidth]{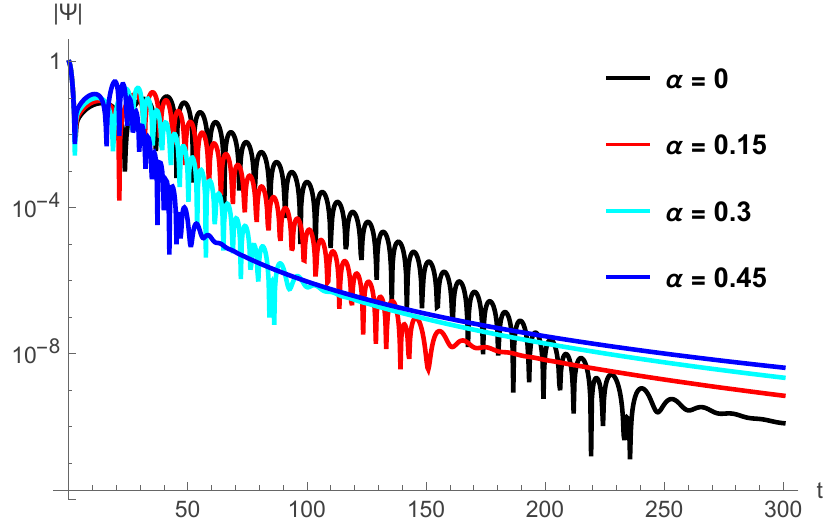}
    \end{minipage}
    \begin{minipage}[b]{0.32\textwidth}
        \centering
        \includegraphics[width=\textwidth]{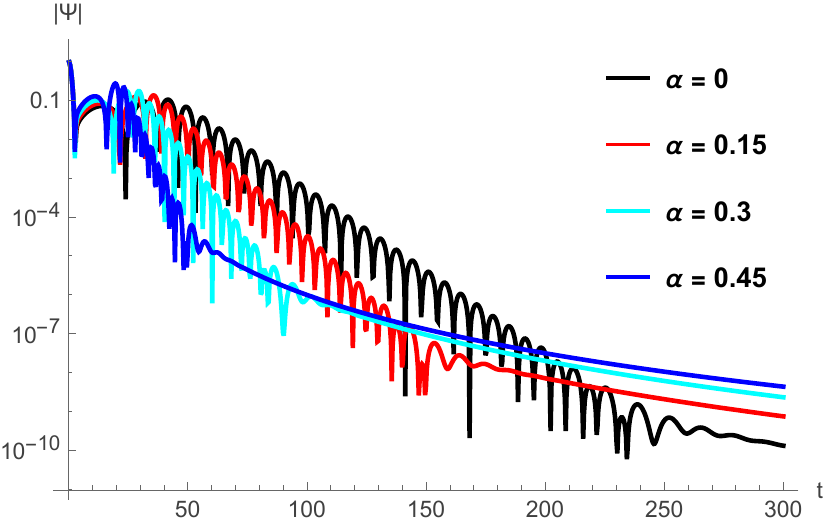}
    \end{minipage}%
    \begin{minipage}[b]{0.32\textwidth}
        \centering
        \includegraphics[width=\textwidth]{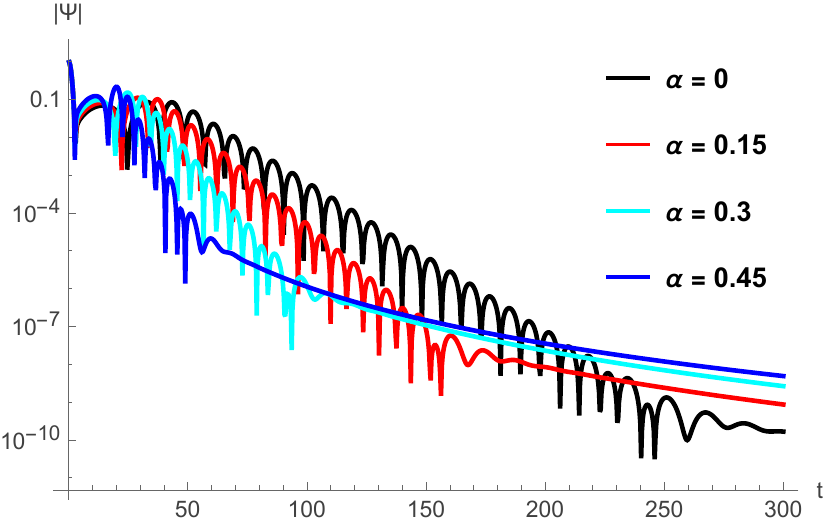}
    \end{minipage}
 \caption{Time-domain profiles under different perturbations.From left to right: scalar field, electromagnetic field, and axial gravitational field, with $\lambda = 0.005, l=2$.}
    \label{fig:8}
\end{figure*}

\begin{figure*}[t!]
    \begin{minipage}[b]{0.32\textwidth}
        \centering
        \includegraphics[width=\textwidth]{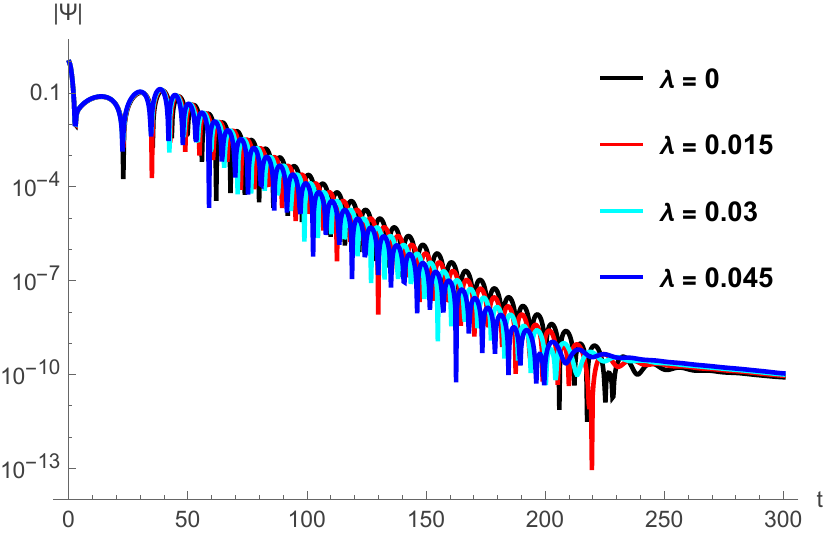}
    \end{minipage}
    \begin{minipage}[b]{0.32\textwidth}
        \centering
        \includegraphics[width=\textwidth]{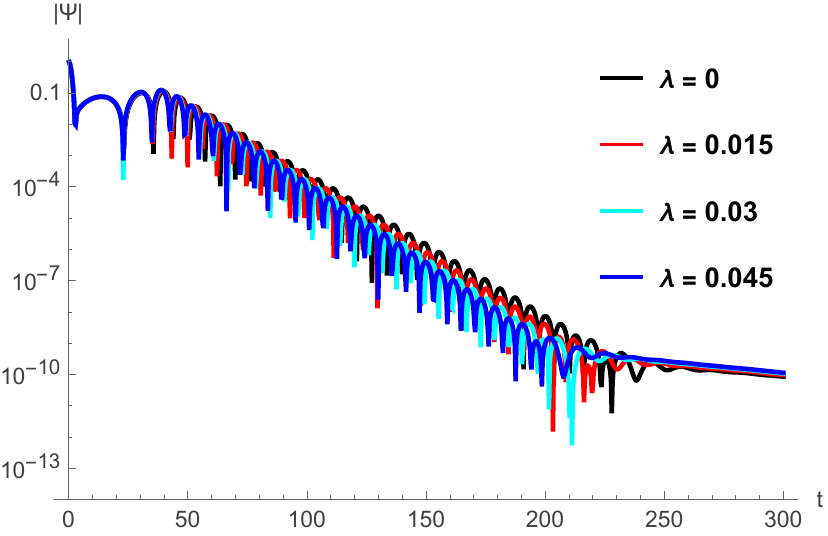}
    \end{minipage}%
    \begin{minipage}[b]{0.32\textwidth}
        \centering
        \includegraphics[width=\textwidth]{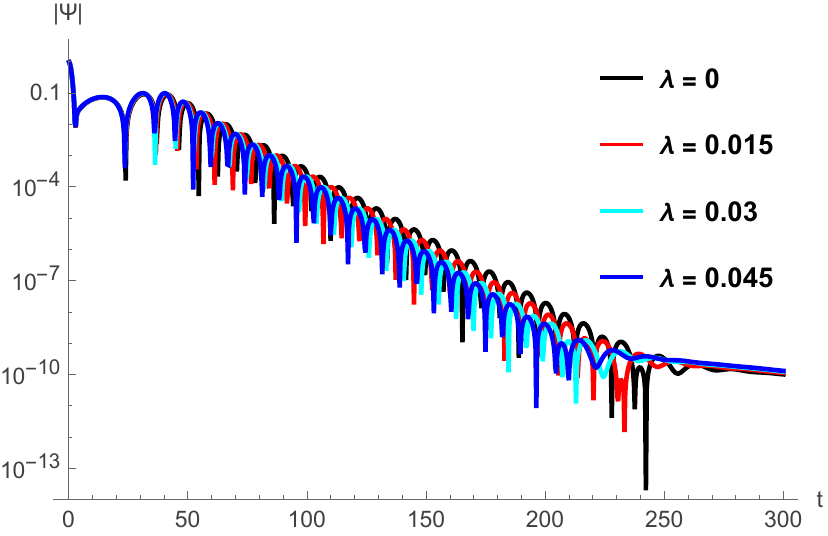}
    \end{minipage}
    \caption{Time-domain profiles under different perturbations.From left to right: scalar field, electromagnetic field, and axial gravitational field, with $\alpha = 0.05, l=2 $.}
    \label{fig:9}
\end{figure*}

When the effective potential exhibits a single-peak structure and vanishes at spatial infinity, the system can be approximated as a one-dimensional barrier scattering problem. In this case, the WKB semi-classical method can be employed to determine the complex frequencies satisfying the quasinormal-mode boundary conditions. The basic idea of this method is to construct a local approximation near the peak of the potential and match the asymptotic solutions on both sides of the barrier, thereby yielding a discrete eigenvalue condition. To improve accuracy, the WKB method has been extended to higher orders; in particular, the sixth-order approximation proposed by Konoplya includes higher-derivative corrections and significantly reduces the errors associated with lower-order approximations~\cite{Konoplya:2003ii}.

\begin{equation}
\frac{i(\omega^2 - V_0)}{\sqrt{-2V_0''}} - \sum_{i = 2}^{6} \Lambda_i = n + \frac{1}{2}, \ (n = 0, \, 1, \, 2 \dots) 
\label{equ:31}
\end{equation}

In this expression, $\omega$ denotes the complex quasinormal-mode frequency. The quantity $V_0$ represents the value of the effective potential at its maximum, while $V_0''$ is the second derivative of the potential with respect to the tortoise coordinate evaluated at the same point.The symbols $\Lambda_i$ ($i=2,\ldots,6$) correspond to the higher-order WKB correction terms, which depend on higher derivatives of the potential at the peak. The integer $n$ is the overtone number, where $n=0$ denotes the fundamental mode.

The time-domain evolution method numerically solves the wave equation of black hole perturbations directly in the time dimension, simulating the whole evolution of the perturbation field from the initial disturbance to gradual attenuation. The quasi-normal mode frequencies are then extracted from the time-domain signal. Introducing the light-cone coordinates$u := t - r_\ast$ and $v := t + r_\ast$, wave Eq. \ref{equ:25} is rewritten as:

\begin{equation}
-4 \frac{\partial^2 \psi(u,v)}{\partial u \partial v} = V \left( \frac{u - v}{2} \right) \psi(u, v)
\label{equ:32}
\end{equation}

To solve Eq.~\ref{equ:32},  discretize it on a null grid in the $(u,v)$ plane using a finite-difference scheme. Considering a null rectangle with vertices denoted by $N$, $W$, $E$, and $S$, corresponding to the points $(u+\Delta u, v+\Delta v)$, $(u+\Delta u, v)$, $(u, v+\Delta v)$, and $(u, v)$, respectively, the value of the field at the future point $N$ can be obtained from the known values at the other three points. This leads to the following  formula~\cite{Abdalla:2005hu}:

\begin{equation}
\psi(N)=\psi(W)+\psi(E)-\psi(S)
-\frac{\Delta u\,\Delta v}{8}\,V(r)\,[\psi(W)+\psi(E)]
\label{equ:33}
\end{equation}

\begin{equation}
\psi(\mu = u_0, \, v) = A \exp \left[ - \frac{(v - v_0)^2}{\sigma^2} \right]
\label{equ:34}
\end{equation}

To initiate the numerical evolution, appropriate initial conditions must be specified on the two null segments $u=u_0$ and $v=v_0$. Following the standard choice in time-domain analyses, we prescribe a Gaussian pulse on one null surface and set the field to zero on the other:

$v_0$, $A$, and $\sigma$ represent the central position of the Gaussian pulse, its amplitude, and its standard deviation, respectively.Substituting the boundary condition Eq.~\ref{equ:34} into Eq.~\ref{equ:33}, the field values at all grid points can be obtained numerically. The results displayed in Fig.~\ref{fig:8} and Fig.~\ref{fig:9} indicate that, following a perturbation, the black hole gradually relaxes toward a stable equilibrium configuration. Among the three types of perturbations, the axial gravitational mode decays the slowest, whereas the scalar and electromagnetic modes exhibit slight differences in their damping behaviors. Increasing either the Lorentz parameter $\alpha$ or the dark matter parameter  
$\lambda$ leads to a shorter duration of the quasinormal ringing.

The time-domain evolution method provides only the dynamical behavior of the perturbation field as a function of time. To extract the quasinormal mode frequencies, an additional analysis is required. In this work, we employ the Prony method to determine the complex frequencies from the time-domain signal. Considering the time interval starting from $t=t_0$ and ending at $t=t_0+Nh$, where $N$ is an integer and $h$ is the sampling time step, the waveform at a fixed position $r_*$ can be expanded as:

\begin{equation}
\psi(t) \simeq \sum_{i = 1}^{p} C_i e^{-i \omega_i t}
\label{equ:35}
\end{equation}

\begin{table*}[t!]
\caption{QNMs frequencies of a black hole for different values of $\lambda$ and $\alpha$, obtained using the sixth-order WKB method and the Prony method, l=2.}
\label{tab:2}      
\begin{tabular}{lcccccc}
\hline
     & \multicolumn{2}{c}{Scalar}   & \multicolumn{2}{c}{Electromagnetic}  & \multicolumn{2}{c}{Axial gravitational}     \\
\hline
\multicolumn{1}{c}{$
$} & \multicolumn{1}{c}{WKB} & \multicolumn{1}{c}{Prony} & \multicolumn{1}{c}{WKB} & \multicolumn{1}{c}{Prony} & \multicolumn{1}{c}{WKB} & \multicolumn{1}{c}{Prony} \\
\hline
$\alpha$ & \multicolumn{6}{c}{$\lambda=0.005$}\\
\hline
0 & 0.491504-0.098459i & 0.492195-0.098005i & 0.464985-0.096667i & 0.465581-0.0962636i & 0.37938-0.090457i & 0.379911-0.090241i \\
0.1 & 0.577109-0.121593i & 0.578251-0.120806i & 0.542626-0.119141i & 0.543597-0.118448i & 0.430297-0.110665i & 0.431243-0.110288i \\
0.15 & 0.629708-0.136347i & 0.631212-0.135283i & 0.589947-0.133438i & 0.591216-0.13251i & 0.459706-0.123483i & 0.460998-0.122928i \\
0.2 & 0.690816-0.153958i & 0.692831-0.152495i & 0.644571-0.150473i & 0.646256-0.149207i & 0.492147-0.138804i & 0.493922-0.137898i \\
0.25 & 0.762489-0.175218i & 0.765247-0.17316i & 0.708177-0.170993i & 0.710459-0.169229i & 0.527955-0.157473i & 0.530354-0.155831i \\
0.3 & 0.847474-0.201207i & 0.851339-0.198238i & 0.782978-0.196011i & 0.786141-0.19351i & 0.567543-0.180833i & 0.570597-0.177607i \\
0.35 & 0.949509-0.233441i & 0.95508-0.229031i & 0.871939-0.226969i & 0.87644-0.223277i & 0.611557-0.211031i & 0.614824-0.204521i \\
0.4 & 1.07379-0.273741i & 1.08209-0.267314i & 0.979115-0.265877i & 0.985724-0.260289i & 0.661195-0.251276i & 0.66291-0.238566i \\
0.45 & 1.22776-0.326368i & 1.24063-0.315511i & 1.11019-0.315725i & 1.12026-0.306941i & 0.718547-0.305452i & 0.714099- 0.283096i\\
\hline
$\lambda$ & \multicolumn{6}{c}{$\alpha=0.05$}\\
\hline
0 & 0.523028-0.107247i & 0.523868-0.106682i & 0.49342-0.105202i & 0.49414-0.104702i & 0.397431-0.098102i & 0.398103-0.097839i \\
0.01 & 0.53818-0.110638i & 0.539096-0.110021i & 0.50761-0.108515i & 0.508395-0.107969i & 0.408512-0.101163i & 0.409239-0.100871i \\
0.015 & 0.544315-0.112048i & 0.545264-0.111408i & 0.513342-0.10989i & 0.514155-0.109324i & 0.412943-0.102431i & 0.413695-0.102127i \\
0.02 & 0.550072-0.113386i & 0.551053-0.112724i & 0.518716-0.111194i & 0.519556-0.110609i & 0.41708-0.10363i & 0.417855-0.103315i \\
0.025 & 0.555555-0.114672i & 0.556566-0.113989i & 0.523829-0.112447i & 0.524695-0.111844i & 0.420999-0.105899i & 0.421797-0.104457i \\
0.03 & 0.560821-0.115919i & 0.561862-0.115215i & 0.528736-0.113661i & 0.529628-0.11304i & 0.424747-0.105899i & 0.425568-0.105561i \\
0.035 & 0.565909-0.117133i & 0.56698-0.116409i & 0.533473-0.114844i & 0.534389-0.114204i & 0.428351-0.106986i & 0.429195-0.106636i \\
0.04 & 0.570844-0.118321i & 0.571944-0.117576i & 0.538064-0.116081i & 0.539005-0.115342i & 0.431832-0.108047i & 0.432699-0.107686i \\
0.045 & 0.575646-0.119487i & 0.576774-0.118721i & 0.542527-0.117134i & 0.543493-0.116458i & 0.435205-0.109087i & 0.436095-0.108714i \\
\hline
\label{tab:2} 
\end{tabular}
\end{table*}

\begin{table*}[t!]
\caption{QNMs frequencies and relative errors obtained using the 6th-order WKB method and the Lyapunov exponent }
\label{tab:3}
\centering
\begin{tabular}{c c
c c c
c c c
c c c}
\toprule

& & \multicolumn{3}{c}{Scalar}
& \multicolumn{3}{c}{Electromagnetic}
& \multicolumn{3}{c}{Axial gravitational} \\
\hline

$l$ & $\omega_s$ 
& WKB & $\Delta R\%$ & $\Delta I\%$
& WKB & $\Delta R\%$ & $\Delta I\%$
& WKB & $\Delta R\%$ & $\Delta I\%$ \\
\midrule
\multicolumn{11}{c}{$\lambda = 0.005$  $\alpha= 0.05$}\\
\hline
2 & 0.433889-0.108472i & 0.531836-0.113615i & 25.9177 & 4.5267 & 0.501765-0.111829i & 15.8235 & 3.0019 & 0.404757-0.10602i & 4.3510 & 2.2605 \\
5 & 1.08472-0.108472i & 1.16312-0.109555i & 10.1523 & 0.98885 & 1.14958-0.109144i & 8.1473 & 0.6157 & 1.10885-0.107856i & 4.7305 & 0.5679 \\
10 & 2.16945-0.108472i & 2.21826-0.10877i & 5.0392 & 0.2740 & 2.21118-0.108656i & 4.4926 & 0.1693 & 2.18987-0.108307i & 3.5632 & 0.1521 \\
20 & 4.33889-0.108472i & 4.32969-0.108551i & 2.5099 & 0.0719 & 4.32607-0.10852i & 2.3668 & 0.0442 & 4.31519-0.108429i & 2.1206 & 0.0398 \\
30 & 6.50834-0.108472i & 6.44139-0.108508i & 1.6712 & 0.0332 & 6.43895-0.108494i & 1.6065 & 0.0203 & 6.43165-0.108453i & 1.4948 & 0.0175 \\
 40 & 8.67779-0.108472i & 8.55315-0.108492i & 1.2525 & 0.0184 & 8.55132-0.108485i & 1.2158 & 0.0120 & 8.54582-0.108461i & 1.1523 & 0.0101 \\
 50 & 10.8472-0.108472i & 10.665-0.108485i & 1.0020 & 0.0120 & 10.6635-0.10848i & 0.9781 & 0.0074 & 10.6591-0.108465i & 0.9372 & 0.0065 \\
\hline
\multicolumn{11}{c}{$\lambda = 0.045$  $\alpha= 0.45$}\\
\hline
2 & 1.02488-0.34965i & 1.31432-0.352656i & 28.2414 & 0.8524 & 1.18664-0.340867i & 13.6318 & 2.5767 & 0.762298-0.326017i & 34.4461 & 7.2490 \\
 5 & 2.56222-0.34965i & 2.83357-0.350388i & 10.5913 & 0.2106 & 2.77606-0.347974i & 7.7037 & 0.4816 & 2.59898-0.340211i & 1.4183 & 2.7748 \\
 10 & 5.12441-0.34965i & 5.38857-0.349859i & 5.1549 & 0.0597 & 5.35849-0.349197i & 4.3684 & 0.1297 & 5.28761-0.34717i & 2.7176 & 0.7143 \\
 20 & 10.2488-0.34965i & 10.5091-0.349705i & 2.5398 & 0.0157 & 10.4937-0.349532i & 2.3338 & 0.0338 & 10.4474-0.349008i & 1.9010 & 0.1839 \\
 30 & 15.3732-0.34965i & 15.6322-0.349675i & 1.6848 & 0.0071 & 15.6218-0.349597i & 1.5914 & 0.0152 & 15.5908-0.349361i & 1.3957 & 0.0827 \\
 40 & 20.4976-0.34965i & 20.7559-0.349664i & 1.2601 & 0.0040 & 20.7481-0.34962i & 1.2073 & 0.0086 & 20.7247-0.349486i & 1.0958 & 0.0469 \\
 50 & 25.622-0.34965i & 25.8799-0.349659i & 1.0066 & 0.0026 & 25.8737-0.349631i & 0.9728 & 0.0054 & 25.8549-0.349545i & 0.9008 & 0.0300 \\
\hline
\end{tabular}
\label{tab:3} 
\end{table*}

In Table~\ref{tab:2}, we present the quasinormal mode frequencies of the black hole under scalar, electromagnetic, and gravitational perturbations for various parameter values. It can be clearly observed that, as the Lorentz parameter $\alpha$ or the dark matter parameter $\lambda$ increases, the real part of the quasinormal frequencies exhibits an overall increasing trend. This behavior is consistent with the increase of the peak height of the effective potential with the parameters. For perturbations of different spins, the scalar field has the largest real frequency, followed by the electromagnetic field, while the gravitational field has the smallest. This ordering corresponds to the relative heights of the effective potential barriers, namely, the scalar field possesses the highest barrier, the electromagnetic field the next, and the gravitational field the lowest.

On the other hand, the absolute value of the imaginary part of the quasinormal frequencies also increases with increasing $\alpha$ or $\lambda$, indicating a faster decay of the perturbations. This trend is consistent with the narrowing of the effective potential barrier, where the influence of $\alpha$ on the barrier structure is more significant, while that of $\lambda$ is relatively weaker. Meanwhile, this result agrees with the time-domain profiles, which show that the duration of the quasinormal ringing becomes shorter as the parameters increase.

Since quasinormal modes are characteristic oscillations determined by the background spacetime geometry and are independent of the specific initial form of the perturbation, these results provide important theoretical support for constraining the Kalb--Ramond field and dark matter parameters through gravitational-wave observations.

In Section \ref{sec:3}, we discussed the observational characteristics of the black hole shadow, photon rings, and the accretion disk; in this section, we further investigate the quasinormal modes generated by perturbations of the black hole in this background. Although these physical phenomena appear to belong to different observational windows, they are essentially governed by the spacetime geometry in the vicinity of the black hole. In particular, in the geometric optics limit, the dominant behavior of quasinormal modes of some black holes is closely related to the dynamical properties of the unstable photon orbit, implying that the shadow size, the photon ring structure, and the spectrum of the ringdown signal are intrinsically connected.
To describe this connection more accurately, one typically considers the limit of large angular quantum number $l \gg 1$, namely the eikonal limit. In this limit, the oscillation frequency and the decay rate of the quasinormal modes are determined respectively by the angular velocity of the orbit and the Lyapunov exponent characterizing its instability~\cite{Cardoso:2008bp,Stefanov:2010xz}:

\begin{equation}
\omega_{S} = \omega_{R} - i\,\omega_{I}
= \Omega\,l - i\,\lambda \left( n + \tfrac{1}{2} \right)
\label{equ:36}
\end{equation}

\begin{equation}
\Omega = \frac{\sqrt{f\!\left(r_{\mathrm{ph}}\right)}}{r_{\mathrm{ph}}}
\label{equ:37}
\end{equation}

\begin{equation}
\lambda
= \sqrt{
\frac{\bigl(2 f(r_{\mathrm{ph}})
- r_{\mathrm{ph}}^{2} f''(r_{\mathrm{ph}})\bigr)\, f(r_{\mathrm{ph}})}
{2\, r_{\mathrm{ph}}^{2}}
}\
\label{equ:38}
\end{equation}

The numerical results in Table~\ref{tab:3} clearly support this picture. As the multipole number $l$ increases, the real part of the frequency grows approximately linearly with $l$, while the imaginary part approaches a constant value. Meanwhile, the frequencies obtained via the 6th-order WKB method progressively converge to those predicted by the eikonal approximation, with rapidly decreasing relative errors. This behavior is consistently observed for scalar, electromagnetic, and axial gravitational perturbations.
These results indicate that, within the validity of the eikonal (WKB) approximation, the high-$l$ quasinormal modes are predominantly governed by the geometry near the photon sphere. Since the same unstable photon orbit also determines the shadow boundary and photon-ring structure, these observables provide complementary probes of the strong-field region around the black hole.

We emphasize, however, that this correspondence is not universal. In more general situations, such as deviations from standard WKB conditions or the presence of non-eikonal modes, the relation between quasinormal modes and null geodesic properties may break down~\cite{Konoplya:2017wot,Konoplya:2022gjp}.

\section{Conclusions}
\label{sec:5}

This paper systematically investigates the optical observational features and ringdown dynamics of a static spherically symmetric black hole in the background of a coupled KR field and PFDM, revealing how the model parameters $\alpha$ and $\lambda$ affect the physical properties in the strong-gravity region. The main results are summarized as follows:

The KR field and PFDM mainly modify the spacetime structure near the event horizon. The event horizon radius $r_{\rm h}$, the photon-sphere radius $r_{\rm ph}$, the shadow radius $b_{\rm c}$, and the innermost stable circular orbit radius $r_{\rm isco}$ all decrease monotonically with increasing $\alpha$ or $\lambda$, indicating a significant ``compressing'' effect of dark matter and the KR field on the black hole spacetime. As $\alpha$ or $\lambda$ increases, the peak observed intensity of the thin accretion disk shifts toward smaller impact parameters.

The effective potentials corresponding to perturbations of different spin fields all exhibit a single-peak barrier structure, with the scalar field having the highest barrier and the axial gravitational field the lowest. As $\alpha$ or $\lambda$ increases, the barrier height increases, while its width decreases with increasing $\alpha$, directly affecting the propagation and damping of perturbations. Both the oscillation frequency (real part $\omega_{\rm Re}$) and the damping rate (absolute value of the imaginary part $|\omega_{\rm Im}|$) of the quasinormal modes increase monotonically with $\alpha$ or $\lambda$. The axial gravitational mode decays the slowest, whereas the scalar mode decays the fastest, consistent with the corresponding potential heights. In the geometric-optics limit of large angular quantum number $l$, the QNM spectrum can be accurately described by the orbital angular velocity and the Lyapunov exponent of the photon sphere, and the relative error between the WKB results and this approximation rapidly converges as $l$ increases.

By jointly analyzing the optical observational signatures and gravitational-wave ringdown signals of this model, this work reveals the combined influence of modified gravity and dark matter on black hole multimessenger signals. The results provide new theoretical support and observable features for confronting related models with astrophysical observations.
\label{sec:4}

\section{Acknowledgements}
This research was  supported by the National Natural Science Foundation of China (Grant No.
12265007).

\bibliographystyle{apsrev4-2}

\end{document}